\UseRawInputEncoding
\documentclass{article}
\usepackage{graphicx} 
\usepackage{caption}
\usepackage{multicol}
\usepackage{amsfonts}
\usepackage{amssymb}
\usepackage{amsmath}
\usepackage{amsthm}
\usepackage{float}
\usepackage{authblk}
\usepackage[dvipsnames]{xcolor}
\usepackage{graphicx}
\usepackage{enumerate}
\usepackage[margin=0.5in]{geometry}
\usepackage{tikz}
\usepackage{siunitx}
\usepackage{hyperref}
\usepackage{url}

\usepackage[normalem]{ulem}
\title{\textbf{On chip high-dimensional entangled photon sources}}
\author[1,2]{Tavshabad Kaur}
\author[1,2]{Daniel Peace}
\author[1,2]{Jacquiline Romero}

\affil[1]{Australian Research Council Centre of Excellence for Engineered Quantum Systems and}
\affil[2]{School of Mathematics and Physics, University of Queensland, Brisbane, QLD 4072, Australia}

\begin{document}
\maketitle
\begin{abstract}

High-dimensional quantum entanglement is an important resource for emerging quantum technologies such as quantum communication and quantum computation. The scalability of metres-long experimental setups limits high-dimensional entanglement in bulk optics. Advancements in quantum technology  hinge on reproducible, and reconfigurable quantum devices --- including photon sources, which are challenging to achieve in a scalable manner using bulk optics. Advances in nanotechnology and CMOS-compatible integration techniques have enabled the generation of entangled photons on millimeter-scale chips, significantly enhancing scalability, stability, replicability, and miniaturization for real-world quantum applications. In recent years we have seen several chip-scale demonstrations with different degrees of freedom including path, frequency-bin, time-bin, and transverse modes, on many material platforms. A complete quantum photonic integrated circuit requires the generation, manipulation, and detection of qudits, involving various active and passive quantum photonic components which further increase the degree of complexity. Here, we review and introduce the nonlinear optical processes that facilitate on-chip high-dimensional entangled photon sources and the currently used material platforms. We discuss a range of current implementations of on-chip high-dimensional entangled photon sources and demonstrated applications. We comment on the current challenges due to the limitations of individual material platforms and present future opportunities in hybrid and heterogeneous integration strategies for the next generation of integrated quantum photonic chips. \\

\end{abstract}

\section{Introduction}

A qubit, the quantum counterpart of a classical bit, can be extended to higher dimensions---qudits. For quantum computation, the higher dimension (\textit{d$>$2}) provides a larger state space for representing and processing quantum information \cite{Wang2020QuditsComputing}. The higher dimensions allow for simultaneous control of multiple operations, a decrease in circuit complexity, simplification of the experimental setup, enhancement of algorithm efficiency, and an increase in computational speed \cite{Wang2020QuditsComputing,Chi2022AProcessor,Du2024Qudit-basedGate}. For quantum communications, qudit-based technology improves security against eavesdroppers, tolerates high bit error rates, and offers better error correction capabilities to improve the integrity of quantum information processing \cite{Cozzolino2019High-DimensionalChallenges,Ecker2019OvercomingDistribution, Sheridan2010SecuritySystems}. \\

High-dimensional quantum information can be encoded on various physical platforms such as Rydberg atoms \cite{Evered2023High-fidelityComputer,Graham2022Multi-qubitComputer,Ghosh2021CreatingAtoms,Zhao2024DissipativeAtoms}, trapped ions \cite{Ringbauer2022AIons,Hrmo2023NativeProcessor}, cold atomic ensembles \cite{Dong2023HighlyMemory,Ding2016High-dimensionalMemories}, superconducting devices \cite{Yurtalan2020ImplementationQutrit,Goss2022High-fidelityCircuits}, spin systems \cite{FernandezdeFuentes2024NavigatingFields}, defects in solid-state devices \cite{Atature2018MaterialTechnologies,Awschalom2018QuantumSpins,Abobeih2019Atomic-scaleSensor}, nuclear magnetic resonance \cite{Dorai2015NovelPHILOSOPHY}, molecular magnets \cite{Moreno-Pineda2021MeasuringTechnologies,Chiesa2024MolecularProcessing,Biard2021IncreasingSpin,Gaita-Arino2019MolecularComputation,Chizzini2022MolecularComputation}, quantum dots \cite{Li2023QuantumReview,Uppu2021Quantum-dot-basedTechnology,Vajner2022QuantumDots,Maring2024APlatform,Halevi2024High-dimensionalTemperature}, and photonic systems \cite{Hu2023ProgressTeleportation,Wang2020IntegratedTechnologies,Kues2019QuantumMicrocombs,Luo2023RecentInternet,Wang2018MultidimensionalOptics,Mahmudlu2023FullyGeneration}. Among these physical platforms, photons are attractive because: they operate as qudits even at room temperature, they interact weakly with the environment, they can be controlled with relatively mature technology, and they can be transmitted across distant nodes more readily compared with matter-based systems \cite{OBrien2009PhotonicTechnologies,Moody2020Chip-scaleGeneration,Rohde2021TheInternet}. Photons facilitate quantum information encoding in different degrees of freedom through continuous-variable (CV) and discrete-variable (DV) approaches, encouraging the transition from theoretical quantum photonic concepts to application-ready technology. In CV quantum information processing (QIP), encoding in quantized amplitude and phase quadratures of electromagnetic fields forms Gaussian states (vacuum states, coherent states, and squeezed states) \cite{Andersen2010Continuous-variableProcessing,Braunstein2005QuantumVariables,Weedbrook2012GaussianInformation}. Discrete-variable QIP, on the other hand, is based on the creation and detection of single photon states encoded in various discrete two-dimensional and high-dimensional degrees of freedom of single photons (qubits and qudits), namely, path \cite{Wang2018MultidimensionalOptics,Llewellyn2020Chip-to-chipSilicon}, polarisation \cite{Zhang2019GenerationSilicon,Lu2007ExperimentalStates}, frequency \cite{Clementi2023ProgrammableDevice,Borghi2023ReconfigurableQudits}, time \cite{Chen2018InvitedWaveguides,Grassani2015Micrometer-scalePhotons} and transverse modes \cite{Feng2019On-chipSource}. The separation between CV and DV has narrowed in recent times with the introduction of hybrid approaches that aim to systematically surmount the inherent limitations of either approach \cite{Andersen2015HybridInformation, Lee2013Near-deterministicQubits}. Regardless of the encoding used, the scalability of photonic systems can be improved by integrating various optical components into a single chip. Among such components, photon sources seek to greatly benefit from enhanced efficiencies offered by coherent pumping of multiple sources in photonic integrated circuits. \\

Discrete-variable encoding requires single photon sources that ideally meet two criteria: (1) deterministic or ``on-demand" generation, and 
(2) indistinguishability. There are currently no sources that can fully satisfy both these requirements. Photon generation can be divided into deterministic and probabilistic approaches. Deterministic single photon sources based on quantum dots, 
color/defect centers, 
single atoms, single ions, single molecules, and atomic ensembles, 
emit single photons following the two-level atomic energy diagram: a photon is produced each time the atom decays from the upper energy state to the lower state~\cite{Smith2019ColourTechnologies,Ripka2018AAtoms,Proppe2023HighlyDots}. When photon emission occurs through a single optical transition, it prevents the generation of more than one photon in the process \cite{Eisaman2011InvitedDetectors,Signorini2020On-chipSources,Lodahl2022APhotons}.
On the other hand, probabilistic sources are based on non-linear parametric processes, generating photon pairs with inherent correlations in time and energy to naturally allow for a ``herald photon" which signals the existence of the other (heralded) photon. While the distinction between a deterministic and a probabilistic source is conceptually clear, this distinction blurs in practice. Deterministic sources become more probabilistic as the coupling efficiencies to other systems (e.g. fibers) decreases, while the success probability of probabilistic sources can be increased by multiplexing many low-probability, but high-fidelity heralded single photons \cite{Meyer-Scott2020Single-photonMultiplexing}. Each of these approaches has advantages and drawbacks. In particular, probabilistic sources are more naturally extended to higher-dimensions compared to deterministic sources. In this review article, we will only address the probabilistic generation of photons. Deterministic photon sources have been reviewed in \cite{Senellart2017High-performanceSources}.\\

Probabilistic quantum light sources based on spontaneous parametric down-conversion (SPDC) or spontaneous four-wave mixing (SFWM) nonlinear optical processes enable the generation of photon pairs in both bulk or integrated platforms. For SPDC a single pump photon is converted into a pair of photons, typically labelled \textit{signal} and \textit{idler} and each roughly half the energy of the pump. Alternatively, in SFWM two pump photons are annihilated to produce the signal-idler photon pair. In both cases the properties of the generated photons (i.e. frequency, polarisation or transverse mode) are such that energy and momentum are conserved. The appeal of SPDC ($\chi^{(2)}$) and SFWM ($\chi^{(3)}$) nonlinear processes for the generation of heralded single-photons \cite{Spring2013On-chipPhotons} and entangled photon pairs on nonlinear material platforms \cite{Kwiat1995NewPairs} lies in their effective operation at room temperature, relatively low preparation and maintenance costs. Nonlinear processes have achieved near-unity levels of indistinguishability, purity, and entanglement fidelity, with pair generation rates approaching GHz \cite{Jin2014EfficientLaser}. These characteristics make SPDC and SFWM sources highly attractive for quantum information applications, such as secure communications \cite{Romero2022EntanglementUp,Moody20222022Photonics}, quantum information processing \cite{Wang2020IntegratedTechnologies,Moody20222022Photonics}, quantum sensing \cite{Degen2017QuantumSensing}, and metrology \cite{Polino2020PhotonicMetrology,Giovannetti2011AdvancesMetrology}. \\

Initial experiments for generating photons were performed in controlled laboratory settings on optical benches on the scale of a few metres. Advances in nanotechnology, development of materials, and fabrication techniques enabled going to the chip scale in order to improve scalability and stability for real-world applications. Traditional down-conversion sources in bulk optics are typically inefficient, requiring high optical intensities or external cavities to enhance the generation rates. However the transition to integrated circuits with the ability to integrate cavity structures, such as microring, microdisk and photonic crystals, provides significant optical confinement and high generation rates across many sources. The rate of generation and detection of photons is highly dependent on the propagation loss, degree of control, and the ability to operate over a broad wavelength range. Quantum information processing in quantum photonic integrated circuits (QPICs) utilizes a combination of active and passive components such as grating couplers, directional couplers, waveguides, ring resonators, Mach-Zehnder interferometers (MZI), multimode interferometers (MMI), multiplexers-demultiplexers, phase shifters, polarization splitters, and delay lines. Collectively, these components influence the various degrees of freedom (DoFs) of photons giving rise to qudit encodings. As the density of components increases, scalability as well as the ability to minimise crosstalk among the components become important.\\ 

Different material platforms and structures have emerged to accommodate the rise in integration levels. Silicon is a leading material that allows for high-density integration, with its high refractive index contrast and established fabrication processes. Silicon is transparent across telecommunication wavelengths and has strong optical nonlinear properties for quantum state generation. Silicon QPICs have progressed from the first CNOT gate and two-photon quantum interference demonstrations \cite{Bonneau2012QuantumCircuits, Silverstone2014On-chipSources} to a fully reconfigurable circuit with over 550 components enabling multidimensional entanglement \cite{Wang2018MultidimensionalOptics}. However, there are are other suitable materials such as silicon nitride, ultra-rich silicon nitride, lithium niobate, III-V semiconductors, silicon carbide, doped silica, and hydex. The limitations of each material prevent any platform from offering the required features for specific quantum applications. However, the field of QPICs has gone a long way in designing and solving the associated problems that come with each material, in some cases with the help of inverse design \cite{Molesky2018InverseNanophotonics} and machine learning \cite{Ma2021DeepStructures, Liu2021TacklingLearning}. All on-chip quantum photonics, from sources to detectors, will inevitably come and enable future technologies like quantum computing and quantum communication.\\

This review discusses various quantum light sources on-chip. Quantum light sources based on SPDC are promising candidates for their high pair generation rate and high signal-to-noise ratio, while the ease in phase-matching is the desirable feature of SFWM due to much closer interacting frequencies. The remainder of this paper is structured as follows. In Section II, we describe the theoretical framework for photon pair generation based on the principles of nonlinear optics and introduce the material platforms of interest. In Section III, we discuss current demonstrations for integrated high-dimensional photon sources across a range of DoFs. In Section IV, we briefly review some relevant applications and methods for system-level integration of emerging quantum photonic platforms. In Section V, we conclude by discussing current and future challenges, and the opportunities for chip-scale applications with high-dimensional entangled states.

\section{Photon Generation with Nonlinear Optics}

Nonlinear optics relates to the study of interactions between light fields that are mediated via a dielectric medium, in particular, the exchange of energy between waves of different frequencies \cite{Saleh1991FundamentalsPhotonics}. These nonlinear processes are divided into two categories: parametric and non-parametric processes. In parametric processes, second-harmonic generation (SHG), sum- or difference-frequency generation (SFG or DFG), third-harmonic generation (THG), four-wave mixing (FWM) and optical parametric amplification (OPA), energy transfer occurs only among the waves. While in non-parametric processes such as stimulated Raman scattering (SRS), stimulated Brillouin scattering (SBS), and two-photon absorption (TPA), photon energy is not conserved with part of the wave energy transferred either from or into the medium. Importantly, in parametric processes the quantum state is maintained which is not the case for non-parametric processes \cite{Sirleto2023AnMaterials}. 
In the presence of an external electric field varying rapidly in time $\textbf{E}(t)$, a dielectric material experiences an induced dielectric polarisation $\textbf{P}(t)$. For weak fields the response is linear such that the dipole moment per unit volume is given by 
\begin{equation}
    \textbf{P}^{(1)}(t)=\epsilon_{0}\chi^{(1)}\textbf{E}(t)
\end{equation}
where $\epsilon_{0}$ is the permittivity of free space and $\chi^{(1)}$ is the linear susceptibility which relates to the linear refractive index ($n_0$) by $n_0 = \sqrt{1+\chi^{(1)}}$. On the other hand, in cases where the electric field strength is strong, such as in a bright optical beam, the material response is nonlinear giving rise to higher order terms described by means of a Taylor expansion of electric field strength, 
\begin{equation}
\begin{split}
\textbf{P} & = \textbf{P}^{(1)}+ \textbf{P}^{(2)}+ \textbf{P}^{(3)}...\\
& =\epsilon_{0}(\chi^{(1)}\textbf{E}+\chi^{(2)}\textbf{E}+\chi^{(3)}\textbf{E}+...)\\
& =\textbf{P}^{(1)}+\textbf{P}_{nonlinear}
\end{split}
\end{equation}
where \textbf{P}$^{(i)}$ and $\chi^{(i)}$ are the i$^{th}$ order polarization and susceptibilities, respectively. Note we have dropped the time dependence for simplicity. The electric fields and dielectric polarisation are given as vectors accounting for the case in which the susceptibilities become tensors of rank $i+1$. The generation of quantum-correlated photon pairs relies on spontaneous parametric processes: spontaneous parametric down-conversion (SPDC) and spontaneous four-wave mixing (SFWM), resulting from $\chi^{(2)}$ and $\chi^{(3)}$ material nonlinearities, respectively \cite{Caspani2017IntegratedOptics,Wang2021IntegratedOptics}. The phase of these dipole oscillations relies on the phases of the incident fields. Hence to enhance the efficiency of the parametric process, the dipoles must act like a phased array - giving rise to a phase matching condition. Generally, phase matching is often a significant challenge (particularly for $\chi^{(2)}$ processes) due to dispersion, limiting the practical applications of the parametric processes.  
However, with dispersion engineering \cite{Lin2007NonlinearApplications}, optical waveguides can achieve phase matching over a broader bandwidth by carefully managing the dispersion properties of signal and idler wavelengths significantly detuned from the pump.

\subsection{Spontaneous Parametric Down-Conversion (SPDC)}

A non-centrosymmetric crystalline material has non-zero even-order susceptibilities owing to asymmetric electronic function \cite{Saleh1991FundamentalsPhotonics}. Therefore, second-order nonlinear optical processes involving three wave components occur when both energy and momentum conservation conditions are satisfied. The $\chi^{(2)}$ nonlinearity of the material facilitates SPDC, where a photon from the strong pump beam ($\omega_{p}$) excites the electron to the virtual level corresponding to $\hbar\omega_{p}$ and then spontaneously decays into two photons: signal ($\hbar\omega_{s}$) and idler ($\hbar\omega_{i}$). Depending on the phase matching condition the generated photons may be either degenerate in which case $\omega_{s} = \omega_{i}$ such that the signal and idler are at half the pump frequency $\omega_{s,i}=1/2\omega_{p}$, or alternatively non-degenerate where $\omega_{s} \neq \omega_{i}$. In either case, the energy and momentum of the entire three-photon process are conserved:
\begin{equation}
\begin{split}
     Energy\hspace{0.1cm}conservation:\hbar\omega_{p}=\hbar\omega_{s}+\hbar\omega_{i},
    \\Momentum\hspace{0.1cm}conservation: \Delta k=k_{s}+k_{i}-k_{p}\approx0,
\end{split}
\end{equation}
where $\Delta$k is the phase mismatch, $\hbar$ is the reduced Planck constant and $k_{p,s,i}$ are magnitudes of wave-vectors of the pump, signal and idler. In SPDC, phase-matching is achieved by two main approaches: using birefringence and quasi-phase-matching (QPM) \cite{Fejer1992Quasi-phase-matchedTolerances,Akbari2013OpticalRanges}. Birefringent phase matching is commonly implemented in bulk crystals as a result of the small phase mismatch experienced. In bulk optics, entangled-photon pairs are generated via SPDC process for various quantum communication protocols, including quantum key distribution and teleportation. The current state-of-art for bulk optics SPDC sources could be found in Refs. \cite{Anwar2020EntangledCrystals, Takeuchi2014RecentApplications, Zhang2021SpontaneousExperiments}. However, owing to advancements at the nano-scale, SPDC waveguide source based on lithium niobate (LN) \cite{Kwon2024Photon-pairConverters}, periodically poled lithium niobate (PPLN) \cite{Ma2020UltrabrightChip,Zhao2020HighWaveguides}, and aluminum-nitride (AlN) \cite{Guo2017ParametricChip} have been demonstrated on-chip for seamless integration of photon sources, single-photon detectors, and waveguide circuits. \\

The SPDC process probabilistically generates a given number of photon pairs such that pair generation rate (PGR) is shown to have following dependence \cite{Wang2021IntegratedOptics},
\begin{equation}
     PGR_{\text {SPDC }} \propto \frac{4 P_p}{9 \varepsilon_0^2 c^2 A_{e f f}}\left(\chi^{(2)} L\right)^2 \operatorname{sinc}^2\left(\frac{\Delta k L}{2}\right)
\end{equation}
where $PGR_{SPDC}$ represents the pair generation rate for SPDC, $P_{p}$ is the pump power, $A_{eff}$ is the mode overlap area, $L$ is the waveguide length, and $\chi^{(2)}$ is the effective value of the second-order nonlinearity tensor. The sinc term accounts for the phase mismatch among the wave components, which limits the effective waveguide length to approximately the coherence length $L_c = \frac{2}{\Delta k}$. Importantly the PGR scales with $1/A_{eff}$ and quadratically with length such that the PGR can be significantly enhanced when moving from bulk to integrated optics as a result of the increased mode confinement and long interaction lengths. Increasing the interaction length $L$, typically comes at the expense of phase matching bandwidth due to the $\operatorname{sinc}^2$ term in which the phase mismatch $\Delta k$ is present. This trade off can be alleviated through dispersion engineering in order to maintain small phase mismatch over longer interaction lengths. 

\subsection{Spontaneous Four Wave-Mixing (SFWM)}

SFWM is involves four wave components, and is a third-order nonlinear process. Using degenerate pumping, two pump photons with the same frequency $\omega_{p}$ generate a pair of photons at frequencies $\omega_{s}$ and $\omega_{i}$. The $\chi^{(3)}$ nonlinearity of the material facilitates SFWM, where two pump photons ($\omega_{p}$) excites the electron to $2\hbar\omega_{p}$ virtual level and then decays by spontaneously emitting photon pairs: signal ($\hbar\omega_{s}$) and idler ($\hbar\omega_{i}$). The energy and momentum of the entire four-photon process are conserved,
 \begin{equation}
 \begin{split}     Energy\hspace{0.1cm}conservation:2\hbar\omega_{p}=\hbar\omega_{s}+\hbar\omega_{i},
    \\Momentum\hspace{0.1cm}conservation: \Delta k=k_{s}+k_{i}-2k_{p}\approx0,
    \end{split}
 \end{equation}
where $\Delta k$ represents the phase mismatch, $\hbar$ is the reduced Planck constant and $k_{p,s,i}$ are magnitudes of wave-vectors of the pump, signal and idler. The strength of $\chi^{(3)}$ nonlinearity of the material and pump beam characteristics strongly affect the energy and momentum conservation principles and the attributes of the single photons. In more detailed treatments, the phase-matching condition includes additional terms related to nonlinear phase accumulation from effects like the Raman scattering, free carrier effect, and Kerr effect\cite{Lin2007Photon-pairPolarization}. Assuming a $\Delta k$ phase mismatch, the  PGR for SFWM can be written as \cite{Foster2006Broad-bandChip,Signorini2019SiliconRange,Signorini2018IntermodalWaveguides,AgrawalG2013NonlinearOptics},
\begin{equation}
    PGR_{\mathrm{SFWM}} \propto\left|\frac{3 \omega_{\mathrm{p}} \chi^{(3)}}{2 n_0^2 \varepsilon_0 c^2 A_{\mathrm{eff}}} P_{\mathrm{p}} L\right|^2 \operatorname{sinc}^2\left(\frac{\Delta k L}{2}\right)=\left|\frac{2 \omega_{\mathrm{p}} n_2}{c A_{\text {eff }}} P_{\mathrm{p}} L\right|^2 \operatorname{sinc}^2\left(\frac{\Delta k L}{2}\right),
\end{equation}
where $\omega_p$ is the pump frequency, $n_{2}$ is the nonlinear refractive index of the waveguide material, \textit{c} is the speed of light, $n_0$ is the refractive index at the pump frequency, and$A_{eff}$ is the mode interaction overlap area. This expression shows the quadratic dependence of photon-pair generation rate on the pump power, nonlinearity of the material, and the inverse of the mode size.  Similar to SPDC, the PGR in SFWM also has quadratic dependence on the waveguide length and the sinc term accounts for the phase mismatch among the wave components. The $\chi^{(3)}$ coefficient, dictating nonlinear susceptibility, is determined by material properties such as atomic arrangement symmetry and atomic properties. In comparison to SPDC, the $\chi^3$ nonlinearity is typically much weaker.\\

Although the PGR of both SPDC and SFWM scale with increasing pump power $P_p$ which comes at the expense of increasing the probability of multiphoton terms. However, theoretical frameworks suggest that multiplexing multiple sources can mitigate this issue, potentially overcoming the challenges associated with increased multiphoton probabilities.


\begin{figure} [hbt!]
    \centering
    \includegraphics[width=0.9\textwidth]{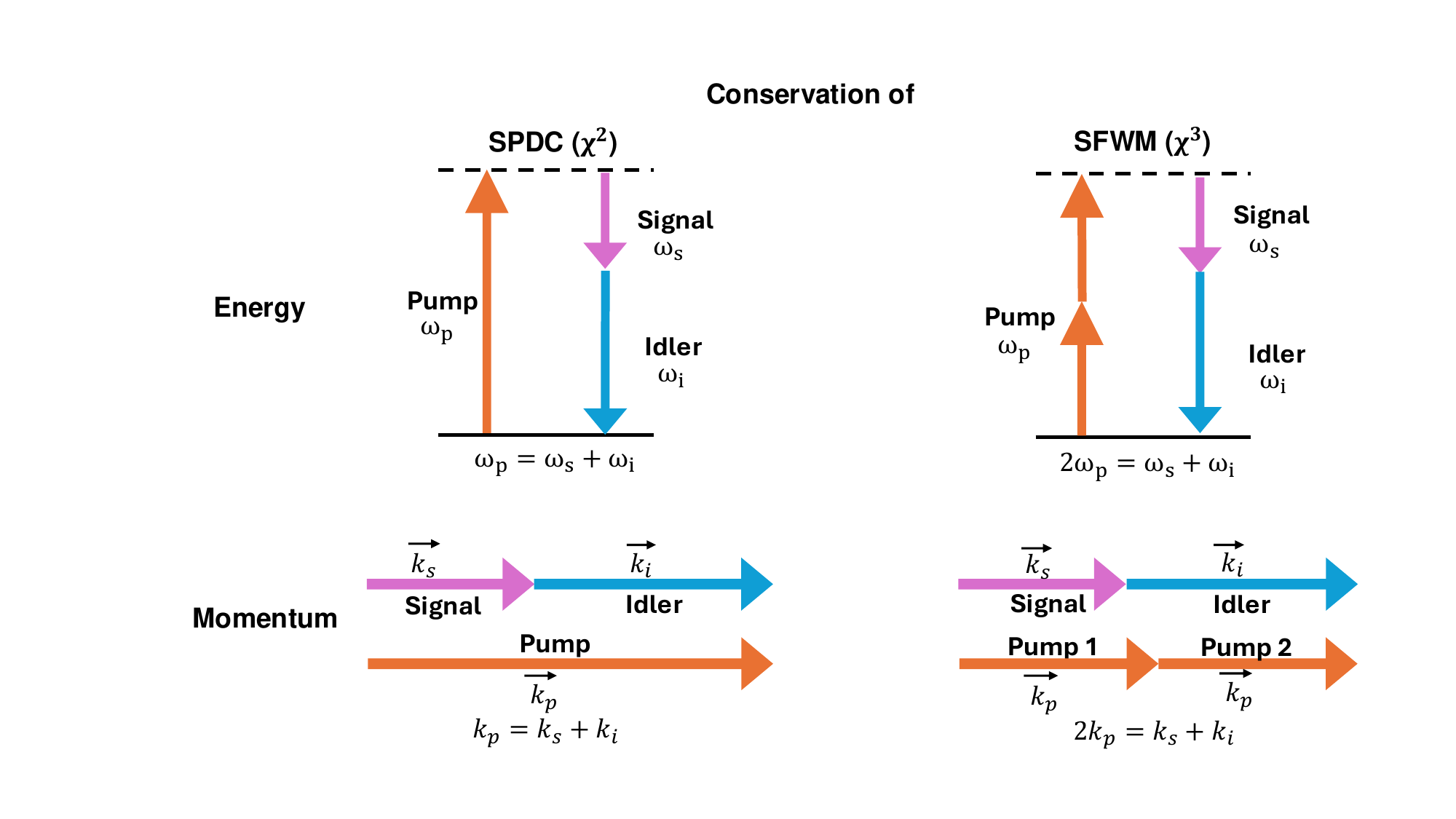}
    \caption{\textbf{Energy and momentum conservation diagrams of spontaneous parametric down-conversion (SPDC) and spontaneous four-wave mixing (SFWM) processes.} In SPDC one pump photon leads to the generation of two correlated photons (signal and idler), while in SFWM, two pump photons lead to the generation of two correlated photons.}
    \label{fig:sfwm_energy_diagram}
\end{figure}

\subsection{Figures of Merit}

\begin{figure} [hbt!]
    \centering
    \includegraphics[width=0.9\textwidth]{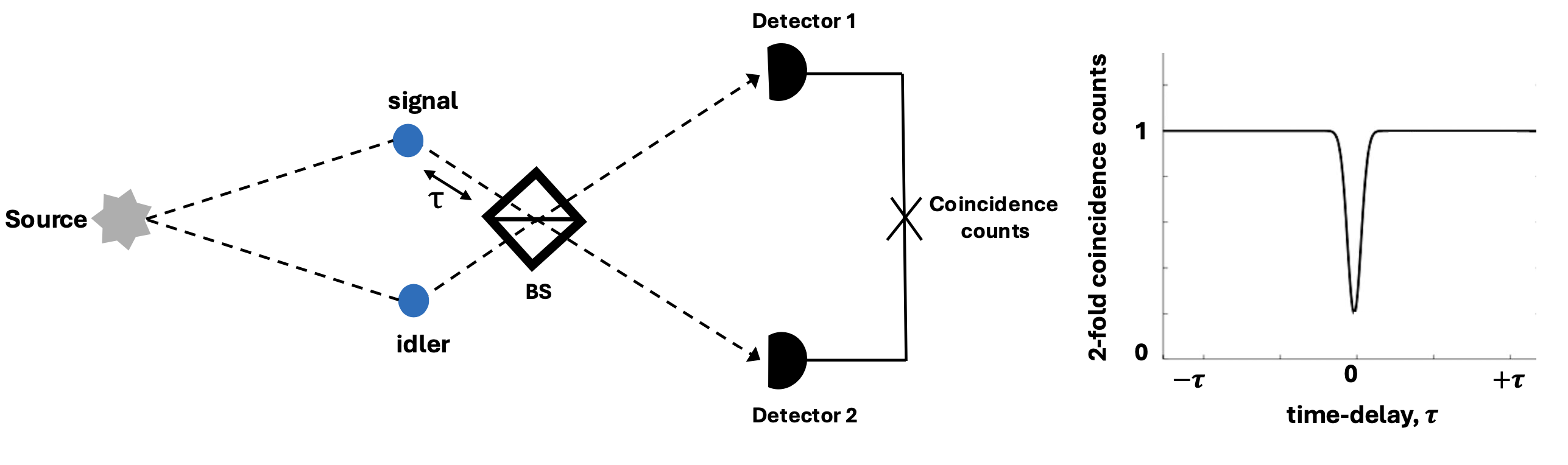}
    \caption{\textbf{Hong-Ou-Mandel experiment (HOM) experiment and 2-fold coincidence measurement plot.} This experiment demonstrates quantum interference between two indistinguishable photons at a beam splitter, resulting in photon bunching, where both photons exit through the same output port, producing a characteristic ``dip" in coincidence measurements (HOM dip), thereby confirming their indistinguishability. Conversely, if the photons are distinguishable, a flat line in coincidence measurements is observed, indicating independent photon exit (anti-bunching).}
    \label{fig:HOM}
\end{figure}

The photon pairs generated by SPDC and SFWM processes are usually characterized by some important parameters to evaluate their performance as photon sources. In this section, we provide a brief introduction to the figures of merit \cite{Faruque2018DevelopingPhotonics, Caspani2017IntegratedOptics,Feng2020ProgressSilicon} used to quantitatively characterize photon pair sources within this review. For a more comprehensive discussion on source properties see Ref.~\cite{Wang2021IntegratedOptics}. 
\begin{itemize}
    \item Single Counting Rate:- In the absence of losses and noise photons, the count rate of signal (idler) photons is identical to the pair generation rate expressed as \cite{Chen2005Two-photon-stateFibers, Sinclair2016EffectWaveguides},
    \begin{equation}
        S_c=A_1(\gamma P_pL)^{2}\frac{\sigma_0}{\sigma_p}I_{sc}
    \end{equation}
where $A_{1}$ is a constant, $\gamma=2 \pi n_2 / \lambda A_{\mathrm{eff}}$ is the nonlinear efficiency, and $I_{sc}$ is a double integral value (the detailed process can be found in \cite{Chen2005Two-photon-stateFibers}). $\sigma_{0}$ is the filtering bandwidth of the signal photon and $\sigma_{p}$ is the pump pulse bandwidth. As indicated in Eq. 7, the single photon counting rate is determined by the square of the pump's peak power. This serves as a key indicator for assessing the impact of noise photons in the system such as spontaneous Raman scattering (SpRS), which scales linearly with pump power. Additionally, the single counting rate is influenced by the ratio of filtering bandwidth to pump bandwidth, with wider filter bandwidth enhancing capture of signal photons, while increased pump bandwidth reduces the interaction time for pump photons. A similar expression can be obtained for CW pump.

    \item Coincidence Counting Rate or Brightness:- Coincidence events refers to the simultaneous detection of the signal and idler photon, which is expressed as, 
    \begin{equation}
        C_c=A_2(\gamma P_pL)^{2}\frac{\sigma_{0}^{2}}{\sigma_p\sqrt{\sigma_{0}^{2}+\sigma_{p}^{2}}}I_{cc}.
    \end{equation}
    Similar to single counting rate, the coincidence counting rate also has a quadratic dependence on the pump power because of the annihilation of two pump photons in the SFWM process. This marks a significant distinction from SPDC photon-pair sources, where the rates of both single and coincidence counting linearly correlate with pump power. The major difference between the single counting rate and the coincidence counting rate depends on the ratio between $\sigma_0$ and $\sigma_p$, which is discussed in detail in Ref.~\cite{Chen2005Two-photon-stateFibers}. 

    \item Coincidence-to-accidental ratio (CAR):- This parameter evaluates the performance of a source in generating photon pairs, accounting for the presence of noise photons. It represents the signal-to-noise ratio of the source, calculated as the ratio between the net coincidence count rate and the accidental coincidences count rate \cite{Caspani2017IntegratedOptics, Takesue2005.5-mFiber}. The net coincidence ($C_{net}$) is the difference of the raw coincidence rate ($C_{raw}$) and the accidental coincidence rate (A).  The accidental coincidence rate includes contributions from the detector dark count rate and from higher order terms of the nonlinear process (which can be minimised by pumping at a suitably low pump power). 
    \begin{equation}
        C_{r a w}=\left(\eta_{\mathrm{c}, \mathrm{s}} \eta_{\mathrm{d}, \mathrm{s}}\right)\left(\eta_{\mathrm{c}, \mathrm{i}} \eta_{\mathrm{d}, \mathrm{i}}\right) r+A \\
        =C_{n e t}+A,
    \end{equation}
    where $\eta_{c/d,s}$, $\eta_{c/d,i}$ and $\eta_{s,i}$ are collection, detection, and total efficiencies of signal and idler photon counting measurements, respectively. Therefore, the net CAR can be written as
    \begin{equation}
        CAR=\frac{C_{net}}{A}
    \end{equation}

    \item Indistinguishability:- It is an important measure representing degree of identicalness among the single-photons emitted from multiple identical photon sources. Two-photon interference effects, experimentally demonstrated by the Hong-Ou-Mandel experiment \cite{Hong1987MeasurementInterference} are often used for characterizing the indistinguishability of the photon, as shown in Figure~\ref{fig:HOM}. When the incoming photons are identical in all degrees of freedom and their wavefunctions completely overlap in time on the beam splitter, a drop in coincidence counts is seen due to their bunching together at the output \cite{Hong1987MeasurementInterference,Faruque2018DevelopingPhotonics}. This HOM dip gives visibility \textbf{(\textit{$V_{HOM}$})} as a quantifier of indistinguishability, which can drop to zero in the case of complete indistinguishability. Experimentally, this is expressed as \cite{Hong1987MeasurementInterference},
    \begin{equation}
        V_{H O M}=\frac{C_{c}(\tau \rightarrow \infty)- C_{c}(\tau=0)}{C_{c}(\tau \rightarrow \infty)},
    \end{equation}
   where $C_{c}$ is the rate of coincidence counts and $\tau$ is the time delay between the two photons. This traditional two photon experiment measuring 2-fold coincidence counts is extended to measure 4-fold coincidence counts using either 4 photons, (2 of which are heralded) or two photons with three beamsplitters in Ref.~\cite{Faruque2018DevelopingPhotonics}. Often on-chip the HOM experiment is performed using the Mach-Zehnder interferometer (MZI) \cite{Rarity1990Two-photonInterferometer,Ou1990ExperimentInterference} where the complete indistinguishability corresponds to 100$\%$ visibility of the interference fringes. 
\end{itemize}

\subsection{Material Platforms}

Engineering photonic devices which facilitate efficient SDPC and SFWM processes in order to yield high-quality entangled photon sources depends greatly on the chosen material platform. In the ideal case for such sources one would choose a material with large nonlinearity, low propagation loss across a wide transparency window and minimal contributions of parasitic effects such as two-photon absorption (TPA), free-carrier absorption (FCA) or photorefractive effects as well as benefits from a mature fabrication process to enable large-scale devices. In practice, no single material is meets all requirements. Recent advancements in photonic material platforms have widened the scope for QPIC implementations, incorporating various active and passive components. Furthermore, compatibility with low-loss optical fibers and the ability for hybrid integration of high-quality lasers and detectors, particularly around telecommunications wavelengths, is crucial for achieving utility scale devices incorporating  on-chip entangled photon sources. Currently, notable platforms for high-dimensional entangled photon pair generation include: Silicon-on-Insulator (SOI), silicon nitride (Si$_{3}$N$_{4}$, SiN), ultra-rich-silicon nitride (USRN), Hydex, and lithium niobate (LiNbO$_3$, LN). In the following paragraphs we will briefly discuss the properties of the aforementioned material platforms as they relate to high-dimensional quantum light sources. \\

Among the mentioned material platforms, SOI is arguably the most mature owing to its CMOS compatibility, driving many recent breakthroughs in integrated photonics \cite{Thomson2016RoadmapPhotonics,Pelucchi2022TheTechnologies,Shekhar2024RoadmappingPhotonics,Leuthold2010NonlinearPhotonics}. Silicon waveguides offer several advantages for on-chip quantum light generation. Firstly, silicon's large nonlinear susceptibility ($\chi^{(3)}\sim 2.8 \times 10^{-18}~\unit{\meter\squared\per\volt\squared}$ \cite{Boyd2008NonlinearOptics}) combined with tight modal confinement, enables efficient SFWM process with modest pump powers \cite{Griffith2015Silicon-chipGeneration,Tsang2008NonlinearWaveguides,Leuthold2010NonlinearPhotonics}. These properties combined with the reliability and scalability of the CMOS foundry processes, has enabled large-scale demonstrations with single chips containing hundreds to thousands of individual components including arrays of coherently pumped sources and large interferometers for programmable quantum state preparation and measurement \cite{Chi2022AProcessor, Bao2023Very-large-scalePhotonics}. Despite these advantages, Si's material properties also lead to several limitations for QPICs. Higher propagation losses resulting from the complex refractive index and increased susceptibility to scattering based on sidewall roughness due to the high refractive index contrast limit overall device transmission - whereas TPA limits photon-pair generation rates at telecommunications wavelengths, and also prohibits the use of shorter wavelengths \cite{Shin2023Photon-pairWaveguide,Liang2004RoleWaveguides,Yin2007ImpactWaveguides,Xiang2020EffectsLasers}. Additionally, due to the absence of an intrinsic Pockels effect, electro-optic (EO) modulators in Si are typically those based plasma dispersion effects, however these suffer from increased losses due to the FCA \cite{Reed2010SiliconModulators}. Alternatively, an effective $\chi^{(2)}$ can be induced via the DC-Kerr effect \cite{Timurdogan2017ElectricWaveguides} which doesn't suffer from increased losses and also operates at cryogenic temperatures \cite{Chakraborty2020CryogenicEffect}. Thermo-optic phase shifters (TOPS) can provide a lower loss alternative based on the thermo-refractive effect, although at the expense of slower modulation speeds \cite{Xie2020Thermally-ReconfigurableCircuits}. As fabrication processes have matured, hybrid integration with other materials naturally possessing an EO effect such as LN or barium titanate (BaTiO$_3$, BTO) has become increasingly popular \cite{Valdez2022110Modulator, Abel2019LargeSilicon}.\\

In addressing the limitations of Si, SiN has emerged as a promising alternative CMOS-compatible material. In comparison to SOI platform, SiN has a lower material loss and larger optical bandgap \cite{Munoz2017SiliconApplications,Kowligy2018TunableWaveguides,Levy2010CMOS-compatibleInterconnects,Kruckel2017OpticalWaveguides} however, it also has a smaller thermo-optic coefficient greatly reducing TOPS efficiency \cite{Arbabi2013MeasurementsResonances}. Plasma-enhanced chemical vapor deposition (PECVD) and low-pressure chemical vapor deposition (LPCVD) techniques have been widely studied for their production high-quality, near stoichiometric Si$_3$N$_4$ layers with low intrinsic losses \cite{Gardeniers1996LPCVDDesign,Kruckel2017OpticalWaveguides}, which combined with the continuous advances in fabrication processes \cite{Liu2021High-yieldCircuits,Pfeiffer2018Ultra-smoothOrigins, Shaw2005FabricationWaveguides,Shao2016Ultra-lowPlatform} has enabled demonstrations of ultra-low propagation ($\leq\unit{\decibel\per\meter}$) losses in waveguide structures \cite{Melchiorri2005PropagationRange,Huang2014CMOSIntegration,Ji2017Ultra-low-lossThreshold,Shaw2005FabricationWaveguides}. Despite its lower third-order nonlinearity in comparison to Si, the larger energy bandgap renders it virtually immune to TPA in the telecom bands \cite{Wang2021IntegratedOptics} enabling the use of high Q-factor cavities to thereby enhancing the overall efficiency and overcoming the reduced nonlinearity.
Improvement in the SiN film deposition~\cite{Bose2024Anneal-freePhotonics} and damascene processes~\cite{Liu2021High-yieldCircuits} have provided thick, crack-free films enabling dispersion engineering at telecom wavelengths which have enhanced various nonlinear processes. An analysis of the dispersion properties of Si$_3$N$_4$ waveguides has been conducted for the effective phase-matching of SFWM process by Hong et.al.\cite{Hong2021DispersionMixing}. Si$_3$N$_4$ resonators have successfully shown the entangled-photon pair generation for various applications such as in quantum sensing and networking \cite{Lu2019Chip-integratedCommunication,Lin2015Si-richGbit/s}. As a compromise between both Si and SiN, USRN has been explored as a method of enhancing the third-order nonlinear coefficient, while still maintaining a wider bandgap. However, the fabrication process of USRN needs significant development to reduce propagation losses in the waveguides \cite{Ng2015ExploringWaveguides,Wang2015SupercontinuumWaveguides,Choi2020CorrelatedWaveguide}. Like Si, the lack of an intrinsic EO effect in SiN has seen the move towards hybrid material platforms in particular with a number of works incorporating LN~\cite{Churaev2023APlatform} and BTO~\cite{Ortmann2019Ultra-Low-PowerSilicon} thin-films, including at demonstrations at cryogenic temperatures~\cite{Eltes2020AnTemperatures}. In a similar fashion, hybrid architectures with SOI or III-V layers have become prevalent in order to augment SiN to overcome some of its limitations \cite{Blumenthal2018SiliconPhotonics,Sharma2020ReviewCircuits,Wilmart2019AApplications,Rahim2017ExpandingCircuits,Moss2013NewOptics}.\\

Lithium niobate, renowned for its strong EO effect and second-order nonlinearity, has been a cornerstone of high-speed modulators for decades~\cite{Zhu2021IntegratedNiobate}.
Similarly, SPDC has long been demonstrated using PPLN waveguides based on Ti in-diffusion~\cite{Martin2010AWavelength} or proton exchange processes~\cite{Tanzilli2001HighlyWaveguide} in bulk. Due to the weak index contrast of these diffused waveguides chip scale demonstrations have largely been restricted to no more than a half-a-dozen components. Along with complexity, the nonlinear efficiency of these devices has also been limited. Developments in wafer bonding over the last decade have enabled commercial availability of high-quality smart-cut wafers with submicrometer thick LN films, spurring significant advances in Lithium Niobate-on-Insulator (LNOI) platform. In particular demonstrations of low propagation losses at visible \cite{Desiatov2019Ultra-low-lossNiobate} and telecom wavelengths \cite{Zhang2017MonolithicResonator} has resulted in significant interest for quantum applications.\\

A relatively large $\chi^{(3)}$ nonlinearity for SFWM facilitates the generation of Kerr combs \cite{Wang2019MonolithicModulation,Yu2020RamanMicroresonators}, while using PPLN waveguides with a large second-order nonlinear coefficient enables efficient SPDC process \cite{Zhao2020HighWaveguides,Zhang2021High-performanceProcesses,Bock2016HighlyWaveguide,Ikuta2019Frequency-MultiplexedConfiguration}. SPDC efficiency in LNOI waveguides has typically been plagued by imperfect phase matching due to variations in film thickness and sub-micron domain widths, though recently the demonstrated efficiency has been greatly improved though adaptive poling methods -- achieving close to theoretical efficiency \cite{Chen2024AdaptedWaveguides}. Alternatively, the efficiency of SPDC can be further enhanced by cavity-based structures \cite{Poberaj2012LithiumDevices,Chang2022IntegratedTechnologies,Wang2023QuantumChips}. The enhanced nonlinear efficiency offered by LNOI waveguides has subsequently reduced the limitations of photorefractive (PR) effects at higher optical intensities previously often found in bulk of or diffused waveguide demonstrations. While LNOI devices can still suffer PR effects, its reduction through annealing and removal of the oxide cladding layer has also recently been studied \cite{Xu2021MitigatingResonators}. The successful generation of entangled photon pairs in LN platform \cite{Xue2021UltrabrightChip,Huang2022High-PerformanceWaveguides,Harper2023HighlyNiobate,Chen2023GenerationWaveguide} is serving as the pathway for the fully reconfigurable QPICs \cite{Jin2014On-chipCircuits,Lomonte2021Single-photonCircuits}. 
Despite these advantages, the high cost of LNOI wafers compared to CMOS-compatible wafers poses a challenge, making it difficult to scale up the fabrication process for high-volume production.\\

Recently, quantum light sources based on CMOS-compatible semiconductor materials have also been demonstrated on silicon carbide (SiC) to host a variety of promising colour centres \cite{Lukin2020IntegratedProspects,Xing2019CMOS-CompatibleOptics,Babin2022FabricationCoherence}. SiC significantly enhances emission from color centers and also exhibits $\chi^{(2)}$ and $\chi^{(3)}$ optical nonlinearities for efficient optical frequency conversion \cite{Lukin20204H-silicon-carbide-on-insulatorPhotonics}. The integration of EOMs, high-Q microresonators, and photonic crystal nanocavities represents a milestone in the advancement of SiC photonic integrated devices \cite{Wang2021High-QPhotonics,Song2019Ultrahigh-QCarbide,Powell2022IntegratedModulator}. Recent studies have shown SiC exhibits an $n_2$ comparable to Si in the 4H polytype \cite{Shi2021PolarizationResonator} enabling demonstration of photon pair source based on SFWM \cite{Rahmouni2024EntangledPlatform}. While the commercial availability of these wafers is still limited, it's expected to improve in coming years. SiCs adoption has been limited in part due to the commercial availability of high-quality SiC wafers through wafer bonding. It's expected this availability will improve in coming years positioning it as a potential future competitor to LN given the improved resistance to photorefractive effects \cite{Wang2021High-QPhotonics} and non-zero $\chi^{(2)}$, $\chi^{(3)}$ and EO coefficients.\\

High-index glass (Hydex), which is a doped fused silica glass, is another CMOS compatible material \cite{Moss2013NewOptics} with refractive index in the range of 1.5 to 1.9 \cite{Wang2021IntegratedOptics}. Hydex integrated waveguides exhibit high nonlinearity and low linear and nonlinear losses, making them promising for nonlinear all-optical signal processing applications \cite{Ferrera2008Low-powerStructures}. Other glasses like 
chalcogenide glasses, which include As$_{2}$S$_{3}$ and As$_{2}$Se$_{3}$, also exhibit high nonlinearities and have attracted considerable attention \cite{Xiong2011GenerationWaveguide,Shiryaev2017RecentPhotonics,Li2014IntegratedDevices}. They have excellent transparency in the mid-IR region and are suitable for photon pair generation via SFWM \cite{Sanghera2009ChalcogenideApplications}. Another group of CMOS-incompatible platforms comprises III-V semiconductor materials \cite{Vyas2022GroupPlatforms}, including GaAs \cite{Stanton2020EfficientWaveguides,Chang2018HeterogeneouslyConversion}, AlGaAs \cite{Mobini2022AlGaAsPhotonics,Steiner2021UltrabrightResonator}, InP \cite{Kumar2019EntangledResonator}, InAs \cite{Delli2019Mid-InfraredSilicon}, AlN \cite{Xiong2012Low-lossProcessing}, and InSb \cite{Jia2018MonolithicPhotonics}. These materials are utilized for generating light across a wide spectrum, from visible to telecommunication wavelengths, owing to their direct bandgap. Among them, AlGaAs is particularly notable for its high third-order nonlinearity and capability to mitigate the effects of TPA through adjustments in the Al concentration. However, despite these advantages, achieving system-level demonstrations for quantum information processing on this platform remains challenging due to its CMOS incompatibility and limited availability of optical components.

\section{High-Dimensional Entangled Photon Sources}

High-dimensional entanglement is advantageous for various quantum technologies such as quantum computation and communication. High-dimensional entanglement enhances the security against potential eavesdropping attempts in quantum key distribution (QKD) \cite{Romero2022EntanglementUp, Erhard2020AdvancesEntanglement}. It also increases the information capacity and improves the error tolerance in measurement-based quantum computing \cite{Romero2022EntanglementUp, Reimer2019High-dimensionalStates}. Such applications need a source of high-dimensional entanglement, which comes for free via the conservation of energy and momentum in nonlinear processes like SFWM and SPDC. High-dimensional entanglement has been often associated with the entanglement of multiple qubits, e.g. the three-particle Greenberger–Horne–Zeilinger (GHZ) states \cite{Zeilinger1997Three-ParticlePairs}. A photon has properties that are naturally amenable to a qudit description, e.g. path, frequency-bin, time-bin, transverse mode, and entanglement in these properties lead to high-dimensional entanglement even with just two photons. The Hilbert space becomes richer as the photon number increases as in multi-photon multi-DOF entanglement.
In practice, the number of modes that can be coherently generated, measured, and controlled restrict the dimensionality of entanglement. \\

There have been plenty of demonstrations of high-dimensional entanglement in bulk optics made possible by the significant $\chi^{(2)}$ nonlinearity of beta-barium borate (BBO)\cite{Malik2016Multi-photonDimensions} and periodically poled potassium titanyl phosphate (ppKTP)\cite{Steinlechner2017DistributionLink} crystals. The advancements in quantum technologies hinge on producing consistent and reproducible quantum devices which is challenging to achieve using bulk optics. Integrated photonics offer advantages in scalability, phase stability, replicability, and miniaturization over bulk optical devices, hence the motivation for putting entangled photon sources on chip. On-chip high-dimensional entanglement via SPDC is achieved using commercially available PPLN \cite{Cabrejo-Ponce2023High-DimensionalDomain} and AlN waveguides \cite{Zhang2023On-chipComb}. High-dimensional entanglement can also be achieved on-chip via SFWM in SOI \cite{Wang2018MultidimensionalOptics} and SiN \cite{Imany201850-GHz-spacedMicroresonator} platforms.\\  

Polarisation is a convenient degree of freedom for investigating entanglement using bulk optics because of the availability of high-brightness entanglement sources.
In 1995, a high-intensity type-II SPDC source of polarisation-entangled photon pairs was made by combining two type-I crystals ("sandwich crystal") \cite{Kwiat1995NewPairs}. 
The first on-chip polarisation-entangled source via SPDC was demonstrated on a waveguide integrated on a PPLN substrate in 2001 \cite{Tanzilli2001HighlyWaveguide}, with a remarkable brightness of \qty[per-mode = symbol]{250}{\mega\hertz\per\milli\watt\per\nano\metre}.
Later works used Bragg-reflection waveguides \cite{Valles2013GenerationWaveguide,Horn2013InherentChip,Zhukovsky2012GenerationChip} and quasi-phase matched waveguides \cite{Takesue2005GenerationCircuit, Sansoni2017AChip,Yoshizawa2003GenerationWaveguides} to improve phase matching and efficiency at the expense of complex waveguide designs. 
The phase matching condition is more readily achieved in SFWM, and polarisation entangled photons have been generated from CMOS-compatible silicon waveguide devices \cite{Lee2008Telecom-BandProcessing, Matsuda2012AChip, Olislager2013Silicon-on-insulatorPhotons}. 
Beyond Si, AlGaAs waveguides exploiting non-vanishing polarisation-mode dispersion, demonstrated polarization-entangled photon pairs via an orthogonally-polarized SFWM process \cite{Kultavewuti2017Polarization-entangledDispersion}. High-dimensional entanglement using polarisation is made possible by having multiple photons, as in a GHZ state. The GHZ state has inspired the development of high-dimensional graph states and cluster states which are relevant to quantum computing \cite{Lu2007ExperimentalStates,Wang201818-QubitFreedom,Zhong201812-PhotonDown-Conversion}.
Multiple polarisation-entangled photon pair sources on a single chip enable the generation of multi-photon entangled states. A four-photon polarization-encoded quantum states created via degenerate SFWM within a spiralled Si waveguide in Sagnac configuration has been demonstrated \cite{Zhang2019GenerationSilicon}, achieving a detection rate of \qty{0.34}{\hertz} at a modest pump power of \qty{600}{\micro\watt} and a fidelity of 0.78$\pm$0.02.\\

\begin{figure}[hbt!]
    \centering
    \includegraphics[width=\textwidth]{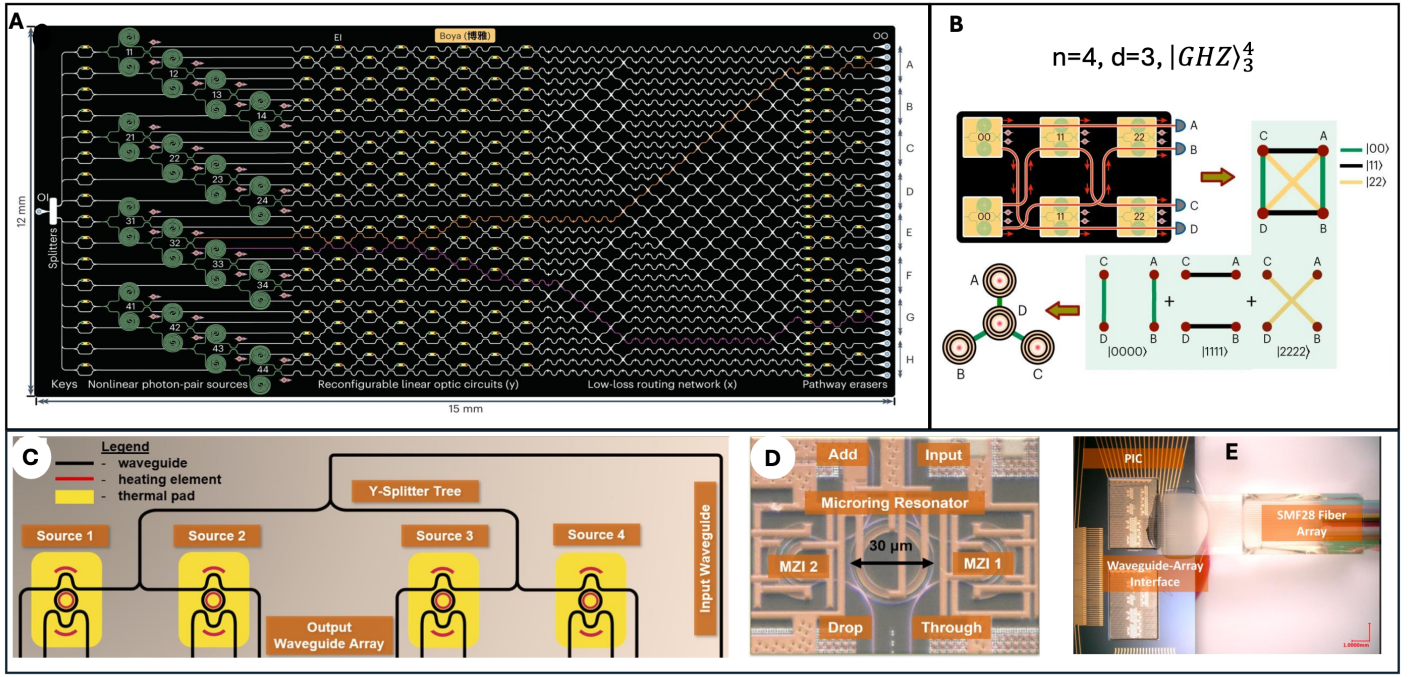}
    \caption{\textbf{High-dimensional sources based on path entanglement in foundry processed chips.} \textbf{(A-B)} A fully programmable quantum device, fabricated on a \qty{200}{\milli\metre} SOI wafer, to generate a synthetic two-dimensional 4$\times$4 lattice array \cite{Bao2023Very-large-scalePhotonics}. \textbf{(A)} Each 12 mm$\times$15 mm `Boya'-graph-based device integrated 32 SFWM photon-pair sources and over 2400 photonic components. The device created 8-vertex graphs, where vertices were single photon pathways from one source to one detector and edges were photon pair sources connecting the vertices. A reconfigurable linear optic circuit in the device altered the connections of the vertices and efficiently controlled the amplitudes and phases of the edges \cite{Bao2023Very-large-scalePhotonics}. \textbf{(B)} This reconfigurable `Boya' device enabled on-chip generation of various multiphoton multidimensional entangled states, for example, four-photon three-dimensional GHZ state \cite{Bao2023Very-large-scalePhotonics}. \textbf{(C-E)} A foundry fabricated silicon PIC \cite{Manfreda2023High-DimensionalChip}. \textbf{(C)} Entangled photon pairs were generated from four interferometrically coupled dual Mach–Zehnder microring resonators acting as photon pair sources. Each source was equipped with heating elements (thermal phase shifters) and thermal pads made of metal \cite{Manfreda2023High-DimensionalChip}. \textbf{(D)} A microscopic image of a single photon pair source \cite{Manfreda2023High-DimensionalChip}. \textbf{(E)} An image of the PIC with light coupled via SMF-28 fiber array \cite{Manfreda2023High-DimensionalChip}. Panels reproduced with permissions from: \textbf{(A, B)} ref.~\cite{Bao2023Very-large-scalePhotonics} under a Creative Commons Attribution 4.0 International License (\url{http://creativecommons.org/licenses/by/4.0/}); \textbf{(C, D, E)} ref.~\cite{Manfreda2023High-DimensionalChip} 2024 Optica Publishing Group under the terms of the Optica Open Access Publishing Agreement.}
    \label{fig:enter-label}
\end{figure}
Path is another degree of freedom that is convenient to implement on an integrated platform---each waveguide corresponds to a possible path that the photon takes. Hong-Ou-Mandel interference has been demonstrated by photons at 1.5$\mu$m, generated by SFWM in two independent Si wire waveguides \cite{Harada2011IndistinguishableWaveguides}. Indistinguishable photons like these can be fed to complex photonic circuits on-chip to generate high-dimensional entangled states. Reconfigurability of the circuit on chip is crucial to enabling programmable generation and processing of quantum information \cite{Shadbolt2012GeneratingCircuit,Chi2022AProcessor}. Photon sources and reconfigurable elements have been demonstrated with SPDC in LN combined with controllable EO phase shifter achieving $92.9 \pm 0.9 \%$ visibilty across the C- and L-bands \cite{Jin2014On-chipCircuits}. On-chip SFWM and programmable phase-shifters have also been demonstrated in Si achieving a very high visibility of $100.0 \pm 0.4 \%$ \cite{Silverstone2014On-chipSources}. Microring resonators have also been used to enhance the efficiency of SFWM. The combination of microring resonators and a programmable MZIs have enabled the production of N00N states that have a $96\pm 2.1 \%$ visibility \cite{Preble2015On-chipSource}. The microring resonator on this chip had high brightness (1$\times$10$^{5}$ photons/s/mW$^{2}$/GHz) and the measured CAR is greater than 500. The entanglement of path-encoded photons have been characterised via quantum state tomography and violation of a CHSH-Bell inequality in \cite{Silverstone2015QubitChip}. Aside from microring resonators, nanostructured photonic crystal slab waveguides (PCSWs) can also amplify nonlinear interactions by leveraging slow light \cite{Baba2008SlowCrystals,Krauss2008WhyLight}. Generation of photons in PCSWs have been demonstrated in \cite{Matsuda2013Slow-light-enhancedWaveguide,Xiong2011Slow-lightWaveguidec}. Scaling up the high-dimensional entanglement in path is achieved by integrating multiple sources on-chip. A foundry-fabricated chip was used by Manfreda-Schulz et. al.~\cite{Manfreda2023High-DimensionalChip} to demonstrate entanglement of photons from four interferometrically coupled dual Mach–Zehnder microring resonators as photon pair sources. An impressive 15$\times$15-dimensional entanglement has been achieved in a chip that integrates more than 550 photonic components on a single chip, including 16 identical photon-pair sources \cite{Wang2018MultidimensionalOptics}. The circuit generates multidimensional bipartite entangled state across 16 optical modes by coherently pumping 16 photon pair sources. The generated photon pairs are separated by asymmetric MZI filters, routed through crossers for local state manipulation, and coupled off the chip by grating couplers to be detected by superconducting nanowire detectors \cite{Wang2018MultidimensionalOptics}. A further demonstration for a very large scale integration has been done by Bao et.al.\cite{Bao2023Very-large-scalePhotonics}, involving a monolithic integration of 2,446 components on a single chip. The 12 mm$\times$15 mm ‘Boya’-graph-based device combines arrays of integrated 32 SFWM photon-pair sources with other linear optical elements to erase the which-source information and to generate multiphoton, multidimensional graph states.\\

\begin{figure}[hbt!]
    \includegraphics[width=\textwidth]{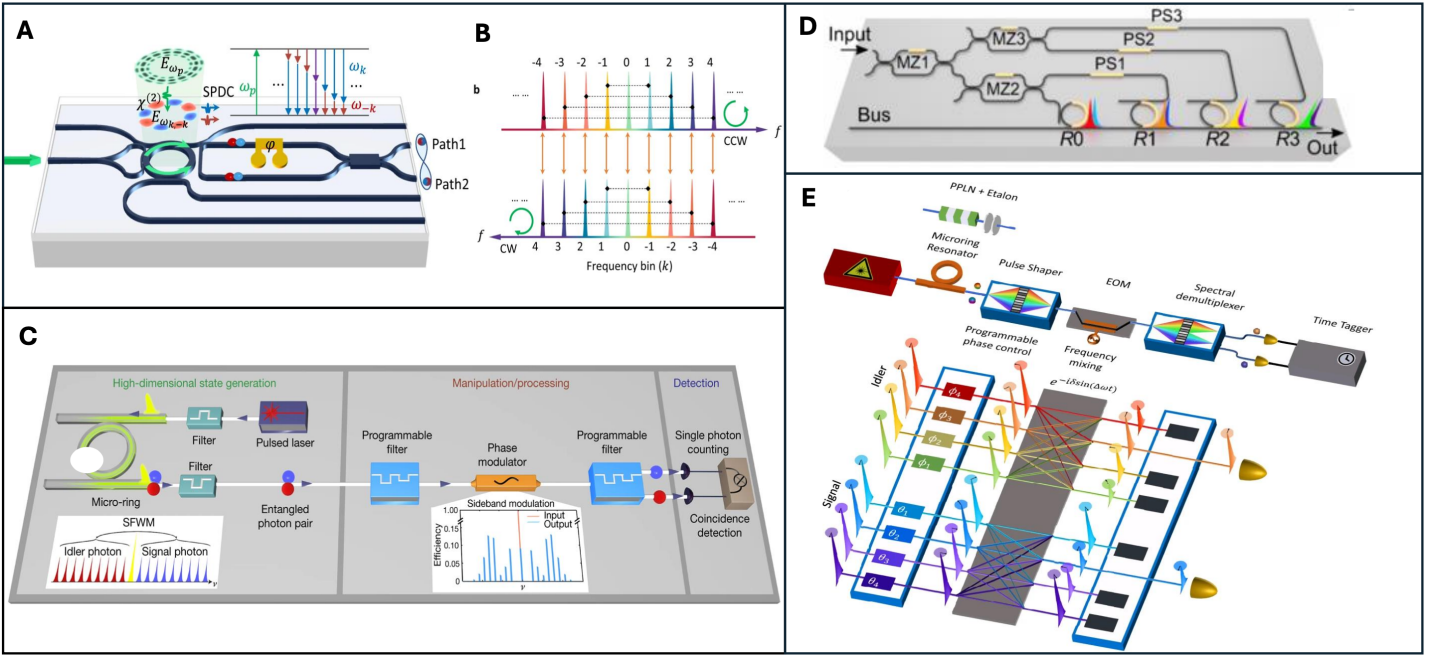}
    \caption{\textbf{High-dimensional frequency-bin entanglement on integrated platforms.} \textbf{(A-B)} Quantum frequency comb (QFC) generation via SPDC \cite{Zhang2023On-chipComb}. \textbf{(A}) A bidirectionally pumped AIN microring resonator based on the Sagnac configuration generated QFC. On-chip quantum interference between parallel comb lines was probed using integrated MZI and TOPS \cite{Zhang2023On-chipComb}. \textbf{(B)} Output QFC state modes in the clockwise (CW) and counter-clockwise (CCW) direction \cite{Zhang2023On-chipComb}. \textbf{(C)} High-dimensional frequency-entangled photon pairs were generated on-chip via SFWM in a nonlinear microring resonator, using a passively mode-locked laser. Programmable filters and phase modulators manipulated the entangled state before detection by single-photon detectors \cite{Kues2017On-chipControl}. \textbf{(D)} Quantum photonic circuit on SOI platform for the realization of reconfigurable frequency-bin entangled qudits. The input light is split into four paths using MZIs (MZ1-MZ3), which excite four microring resonators (R0-R3) to generate signal-idler photon pairs across different frequency bins~\cite{Borghi2023ReconfigurableQudits}. \textbf{(E)} On-chip biphoton frequency generation via SiN microring resonator for upto 8$\times$8-dimensional two-qudit Hilbert space~\cite{Lu2022BayesianMeasurements}. Panels reproduced with permissions from: \textbf{(A, B)} ref.\cite{Zhang2023On-chipComb} under a Creative Commons Attribution 4.0 International License (\url{http://creativecommons.org/licenses/by/4.0/}); \textbf{(C)} ref.~\cite{Kues2017On-chipControl} Springer Nature ; \textbf{(D)} ref.~\cite{Borghi2023ReconfigurableQudits} APS ; \textbf{(E)} ref.~\cite{Lu2022BayesianMeasurements} under a Creative Commons Attribution 4.0 International License (\url{http://creativecommons.org/licenses/by/4.0/}).}
    \label{fig:frequency_encoding}
\end{figure}

Qudits can also be encoded using frequency. Frequency modes that readily travel in an optical fibre offer a practical avenue to scale up high-dimensional entanglement. Quantum frequency combs (QFCs) feature multiple phase-stable frequency modes within a single spatial mode \cite{Kues2019QuantumMicrocombs,KippenbergMicroresonator-BasedCombs}. Quantum information in these quantum frequency modes is a rapidly growing area of research \cite{Chen2013ExperimentalComb,Xie2015HarnessingComb,Cai2017MultimodeCombs,Wang2019MonolithicModulation}. High-dimensional frequency-bin entangled photon pair generation at telecom wavelengths via SPDC in PPLN waveguide, together with coherent control of eleven frequency bins was demonstrated in \cite{Olislager2010Frequency-binPhotons}. The on-chip parallel processing of QFCs utilizing an integrated aluminium nitride (AlN) platform in Sagnac configuration was demonstrated by Zhang et al.~\cite{Zhang2023On-chipComb}, achieving high-visibility quantum interference and high-fidelity state control across all frequency modes. This advancement enables the deterministic separation of photon pairs in QFCs without spectral filtering, demonstrating a high-dimensional HOM interference. QFCs are also generated via SFWM in SOI micro-ring resonators, such as in \cite{Chen2011Frequency-binMicro-resonator} which showed 21-pairwise correlations in frequency bins spanning from \qty{1.3}{\micro\metre} to \qty{1.8}{\micro\metre}. The compatibility of SOI with CMOS processes is advantageous, but the significant TPA at telecom wavelengths poses a fundamental limitation. This motivated studies of other CMOS-compatible platforms, e.g. high-index silica glass and SiN \cite{Moss2013NewOptics}. The generation of high-purity photons with large CAR values rely on narrow spectral bandwidths of the pump in the SFWM process. This requirement was addressed by Kues et.al.~\cite{Kues2017PassivelyWidth} and Roztocki et.al.~\cite{Roztocki2017PracticalCombs} which used a passive mode-locked laser system relying on a nested-cavity configuration. A passive mode-locked laser was used in \cite{Kues2017On-chipControl} for SFWM in a micro-ring resonator in high-index silica glass, showing on-chip generation of entangled qutrit states ($d{=}3$, fidelity=80.9$\%$) and entangled ququart states ($d{=}4$, fidelity=76.6$\%$). This demonstration also required reducing the free spectral range (FSR) of the resonator and programmable filters with a higher frequency resolution. A SiN micro-ring resonator with a FSR of 50 GHz was used in \cite{Imany201850-GHz-spacedMicroresonator} to demonstrate on-chip qubit and qutrit frequency-bin entanglement in a frequency comb consisting of 40 mode pairs.\\

Although frequency modes offer compatibility with telecom networks, quantum interference and the measurement of superposition states is a challenge because these require a nonlinear optical process \cite{Brecht2015PhotonScience}. An active ``frequency beam splitter" which resulted in an interference visibility of 95$\pm 2 \% $ was developed in \cite{Joshi2020Frequency-DomainMicroresonator} by exploiting Bragg-scattering four-wave mixing in an optical fibre. Combined with phase-shifting, the frequency beamsplitter can be the basis of high-fidelity two-photon operations in the frequency domain. While the frequency beamsplitter in \cite{Joshi2020Frequency-DomainMicroresonator} can provide larger frequency shifts from a nonlinear process, it is more common to use EO frequency mixing, which is more convenient but capable of imparting only small frequency shifts. It then becomes important to have tightly spaced frequency bins, while also maintaining high brightness---two competing constraints that cannot be addressed by a single micro-ring resonator because of the inherent trade-off between these two requirements. Being able to program several integrated micro-ring resonators such that they each cover different frequency bins becomes beneficial. In \cite{Borghi2023ReconfigurableQudits}, four identical rings in SOI achieved a high brightness of $0.63 \pm 0.15$ MHz/(mW)$^{2}$ per comb line with a bin spacing of 15 GHz. As shown Fig.~\ref{fig:frequency_encoding}D, each ring is pumped at a different wavelength with mutually coherent pumps to generate entangled states up to a dimension of 16, yielding fidelities above 85$\%$ for maximally entangled Bell states. A similar work using two SOI micro-ring resonators can generate all four of the maximally entangled Bell states achieving a fidelity of 97.5$\%$ and purity close to 100$\%$ \cite{Clementi2023ProgrammableDevice}. In both these works, tuneable MZIs facilitated the manipulation of field intensity and the relative phase for each ring during the SFWM process. The decoupling of the generation rate from the frequency bin spacing will enable more dense integration, with increased number of coherently excited rings leading to more complex quantum states for quantum information processing. The next step in improving the scalability of these on-chip sources is to incorporate the pump laser and the subsequent filtering also on-chip, making these sources less bulky and practical to use outside laboratories. A fully integrated source of frequency-entangled photons was demonstrated in \cite{Mahmudlu2023FullyGeneration}. Their design integrates a laser cavity, a tuneable noise suppression filter ($>$55 dB) utilizing the Vernier effect, and a SiN microring for the generation of entangled photon pairs. The chip achieves a pair generation rate of 8,200 counts s$^{-1}$, with state fidelity of 99$\%$ and interference visibility of 96$\%$. \\

With the advent of advanced frequency-entangled sources, advanced characterisation techniques need to be developed. While joint spectral intensity measurements show the correlations among the frequency-bin pairs, these are insensitive to phase coherence and hence not useful for characterising entanglement. The active frequency mixing required to do projective measurements involves strong filtering of the input quantum state \cite{Kues2017PassivelyWidth}. The alternative, which is to design programmable qudit gates for quantum state tomography inevitably increases in complexity as the dimension increases \cite{Liu202240-userNode}. In \cite{Lu2022BayesianMeasurements}, rather than measuring in the standard informationally complete bases, the complex and random frequency mixing behaviour of EOMs was exploited in conjunction with pulse shaping. The result are randomised operations, which together with the coincidence measurements can be fed to a Bayesian algorithm to obtain the density matrix. The Bayesian approach is numerically more complicated than the standard quantum state tomography, but its applicability to quantum state tomography of a bipartite high-dimensional entangled state generated on a SiN micro-ring resonator has been demonstrated in \cite{Lu2022BayesianMeasurements} (Fig.~\ref{fig:frequency_encoding}E). With this technique, the density matrix of an 8$\times$8-dimensional Hilbert space---the highest dimension to date for frequency bins---was obtained. \\

\begin{figure} [hbt!]
    \centering
    \includegraphics[width=\textwidth]{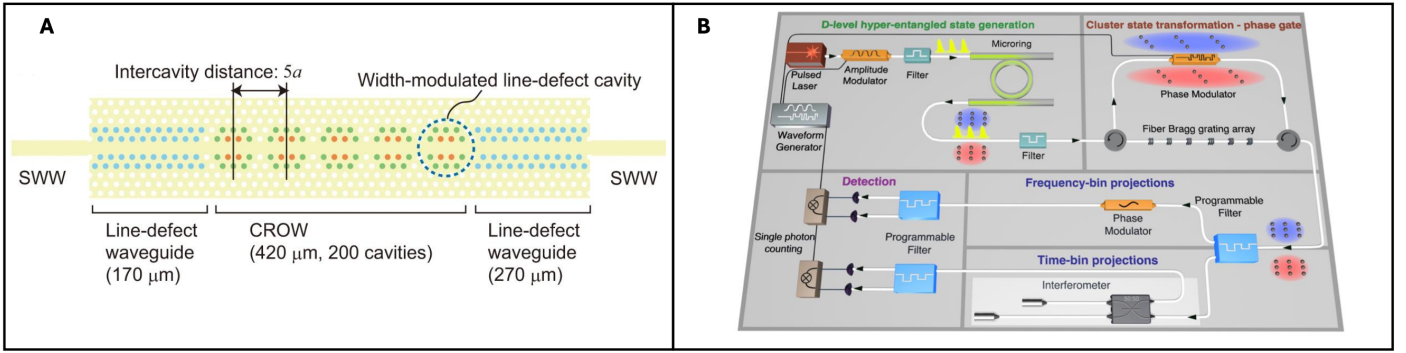}
    \caption{\textbf{High-dimensional time-bin and time-frequency/energy entanglement on integrated platforms.} \textbf{(A)} Demonstration of high-dimensional time-bin entanglement using an on-chip slow light device, a coupled resonator optical waveguide (CROW). The CROW consisted of 200 silicon photonic crystal nanocavities pumped by a coherent pulse train to generate a high-dimensional time-bin entangled state~\cite{Takesue2014EntangledLight}. \textbf{(B)} Experimental setup for generating d-level cluster-states and performing measurement-based quantum computation operations. A nonlinear microring resonator generated signal and idler photons pairs in the superposition of three time and three frequency modes to form a hyper-entangled state~\cite{Reimer2019High-dimensionalStates}. Panels reproduced with permission from: \textbf{(A)} ref.~\cite{Takesue2014EntangledLight} under a Creative Commons Attribution 3.0 Unported License (\url{http://creativecommons.org/licenses/by/3.0/}); \textbf{(B)} ref.~\cite{Reimer2019High-dimensionalStates} Springer Nature.}
    \label{fig:time-bin}
\end{figure}
Temporal degree of freedom is another property of photons that can be entangled. An 11-dimensional time-bin entangled state was generated via SPDC in KNbO$_{3}$ nonlinear crystal, with visibility of 91$\pm$6$\%$ \cite{deRiedmatten2002CreatingLasers}. The footprint of the photon pair generating source was reduced by utilising AlGaAs Bragg-reflection waveguides, which generated high-dimensional time-bin entanglement with a visibility of 94.2$\pm$9$\%$ on a chip \cite{Chen2018InvitedWaveguides}. The first on-chip high-dimensional entanglement using 200 silicon photonic crystal nanocavities was proposed by Takesue et.al.~\cite{Takesue2014EntangledLight} (Fig.~\ref{fig:time-bin}A), demonstrating a high-dimensional time-bin entangled photon source with much smaller footprint. The work by Fang et.al.~\cite{Fang2018On-chipEntanglement}, aimed at achieving high-capacity quantum communication using a silicon nanowire waveguide photon pair source, demonstrated the distribution of time-bin entangled photons independently into 3(time)$\times$14(wavelength) channels. Frequency-bin encoding involves measuring the frequency/wavelength correlations without any information about the time of arrival of photons, while time-bin encoding measures arrival time correlations of photons. Time-frequency encoding combines both, capturing correlations in both the domains. In frequency-bin and time-bin entanglement, the precise frequency measurements increase uncertainty in arrival times and vice versa, making joint time-frequency measurements essential for a complete characterisation of entangled states. Time-energy entanglement is tested with the Franson interferometer by sending pairs of photons through unbalanced interferometers with different path lengths. By measuring the interference fringes resulting from the phase differences between the interferometer arms, nonlocal quantum correlations are verified \cite{Kwiat1993High-visibilityTime,Agne2017ObservationInterference}.
Some examples of experimental demonstrations of time-frequency/ time-energy entanglement can be found in \cite{Maclean2018DirectPairs, Grassani2015Micrometer-scalePhotons,Mazeas2016High-qualityChip,Ma2017Silicon006} 
High-dimensional time-energy entanglement generation on fiber-ppKTP was reported by Cheng et.al.~\cite{Cheng2023High-dimensionalComb} with a visibility of 99.8$\%$. The first experimental realization of qudit cluster states using time-frequency entanglement, was demonstrated by Reimer et.al.~\cite{Reimer2019High-dimensionalStates} via SFWM in a Hydex microring using a series of mode-locked pulses, to perform high-dimensional one-way quantum processing (Fig.~\ref{fig:time-bin}B). \\

\begin{figure} [hbt!]
    \centering
    \includegraphics[width=\textwidth]{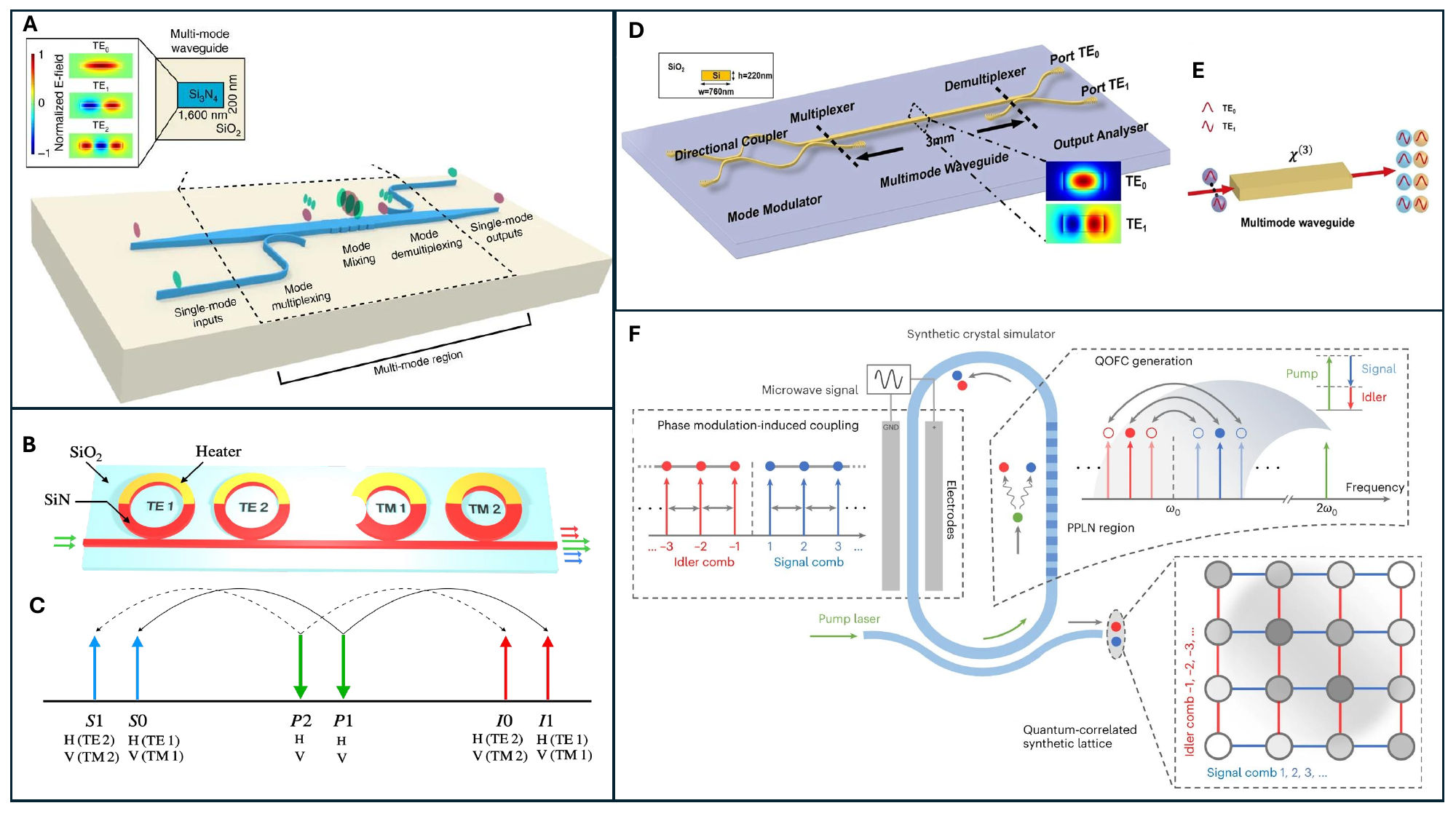}
    \caption{\textbf{High-dimensional transverse mode entanglement, hyper-entanglement and synthetic space on integrated platforms.} \textbf{(A)} On-chip quantum interference between the transverse spatial modes (TE$_0$, TE$_1$, and TE$_2$) within a single SiN multi-mode waveguide. The chip comprised of spatial mode multiplexers (asymmetric directional coupler) and a spatial mode beamsplitter (nanoscale grating) \cite{Mohanty2017QuantumModes}. \textbf{(B-C)} On-chip demonstration of hyper-entangled (polarisation and frequency) photon pairs using a series of four ring resonators \cite{Vendromin2023ProgrammablePairs}. \textbf{(B)} Each SiN ring generated photon pairs by SFWM, with both photons sharing the same polarization – either horizontal (H) or vertical (V) – and generated in two possible frequency-bin pairs within the ring. Each ring was equipped with a heater to fine-tune its resonances \cite{Vendromin2023ProgrammablePairs}. \textbf{(C)} Relative positions of the pump resonances and signal and idler frequency bins \cite{Vendromin2023ProgrammablePairs}. \textbf{(D-E)} On-chip transverse mode encoded photon pair generation \cite{Feng2019On-chipSource}. \textbf{(D)} Schematic of the SOI chip comprising a mode modulator, a multimode waveguide, and an output analyzer for separating the two modes. Entangled photon pairs were generated in a 3 mm long multimode Si waveguide via SFWM process~\cite{Feng2019On-chipSource}. \textbf{(E)} Types of photon pair generation in a multimode waveguide: Intramodal and Intermodal FWM~\cite{Feng2019On-chipSource}. \textbf{(F)} A LN racetrack resonator produced a 2D quantum-correlated synthetic lattice using time–frequency entangled photons generated via SPDC within the synthetic crystal simulator. The resulting QFC was structured into a tight-binding quantum-correlated synthetic crystal through EO modulation~\cite{Javid2023Chip-scaleSpace}. Panels reproduced with permission from: \textbf{(A)} ref.~\cite{Mohanty2017QuantumModes} under a Creative Commons Attribution 4.0 International License (\url{http://creativecommons.org/licenses/by/4.0/}); \textbf{(B, C)} ref.~\cite{Vendromin2023ProgrammablePairs} under CC BY 4.0 Attribution 4.0 International Deed \url{(https://creativecommons.org/licenses/by/4.0/}; \textbf{(D, E)} ref. \cite{Feng2019On-chipSource} under a Creative Commons Attribution 4.0 International License (\url{http://creativecommons.org/licenses/by/4.0/}); \textbf{(F)} ref.~\cite{Javid2023Chip-scaleSpace} Springer Nature.}
    \label{fig:Tmodes-Hyper-entanglement-Synthetic}
\end{figure}

Beyond spectral and temporal degrees of freedom, the transverse modes present in multimode optical waveguides enhances parallelism and scalability compared to single-transverse-mode configurations. Using such spatial modes are reminiscent of optical communication systems that enhance information capacity via spatial multiplexing techniques~\cite{Richardson2013Space-divisionFibres}. Entanglement in transverse DoF can be converted to path and polarization entanglement, offering control over multiple degrees of freedom simultaneously \cite{Feng2016On-chipFreedom}. High-dimensional spatial mode entanglement via type-II SPDC process was demonstrated by Bharadwaj et.al~\cite{Bharadwaj2015GenerationCoupler} using a three-waveguide directional coupler in a PPLN substrate. The width and the height of the three-waveguide coupler \cite{Bharadwaj2015GenerationCoupler} were designed to achieve a two-photon output state occupying three different transverse spatial modes. Mohanty et.al.~\cite{Mohanty2017QuantumModes} designed multimode SiN waveguides supporting three spatial modes (TE$_0$, TE$_1$, TE$_2$) and demonstrated tunable quantum interference between pairs of photons in different transverse spatial modes using a grating structure along the multimode waveguide (Fig.~\ref{fig:Tmodes-Hyper-entanglement-Synthetic}A). The demonstration of intermodal four-wave mixing process in integrated multimode silicon waveguides \cite{Signorini2018IntermodalWaveguides, Signorini2019SiliconRange,Paesani2020Near-idealPhotonics} has laid a foundation for future advancements in silicon photonics. Building on these works, Feng et. al.~\cite{Feng2019On-chipSource} made an advancement by reporting the first-ever on-chip multimode SFWM silicon waveguide photon pair source (Fig~\ref{fig:Tmodes-Hyper-entanglement-Synthetic}B), paving the way towards higher-dimensional quantum systems. Transverse-mode entangled photon pairs were verified across various frequency channels within $\approx{2}$ THz bandwidth, achieving a 0.96$\pm$0.01 high-fidelity Bell state. \\

Besides high-dimensional encoding in any one DoF, the simultaneous entanglement in multiple DoF---hyperentanglement---enhances the quantum information processing by encoding more information per photon. Hyperentangled states increase channel capacity and noise resistance, improving QKD protocols \cite{Hu2021PathwaysNoise,Doda2021QuantumEntanglement,Islam2017PQudits}. The first experimental demonstration of a 36-dimensional quantum system entangled in polarisation, spatial mode and time-energy was done in \cite{Barreiro2005GenerationPairs}, generating hyperentanglement. Hyperentangled photon pairs have since been generated in various configurations, including frequency-polarisation \cite{Lu2023TomographyPhotons,Francesconi2023On-chipStates}, polarisation-energy-time \cite{Huang2022High-PerformanceWaveguides,Xie2015HarnessingComb}, path-frequency \cite{Zhang2023On-chipComb}, polarisation-spatial modes \cite{Kang2014HyperentangledWaveguides}, implemented on integrated platforms through the SPDC process. Hyperentanglement was also shown in Bragg reflection waveguides \cite{Zhukovsky2011ProposalHyperentanglementb} and AlGaAs ridge waveguides \cite{Kang2014HyperentangledWaveguides}. Recent years have seen much exploration of hyperentangled photon pair generation via micro-ring cavities as demonstrated by Suo et al.~\cite{Suo2015GenerationCavity}. In this work \cite{Suo2015GenerationCavity}, a scheme of hyper-entangled polarization and energy-time photon pair generation based on a silicon micro-ring cavity achieving $>94\%$ visibility has been demonstrated. Work by Vendromin et al.~\cite{Vendromin2023ProgrammablePairs} introduced a system comprising four SiN microring resonators on a chip (Fig.~\ref{fig:Tmodes-Hyper-entanglement-Synthetic}B-C), capable of generating polarization and frequency-bin entangled photon pairs. In general, photonic structures are described by physical geometric dimensions such as zero-, one-, two-, and three- dimensional structures. When an additional degree of freedom---synthetic dimension---is combined with the geometrical dimensions, it enables the exploration of higher-dimensional synthetic spaces within simpler, lower-dimensional physical structures \cite{Yuan2018SyntheticPhotonics}. Synthetic space is formed by utilizing various photonic modes such as frequency, OAM, or temporal modes, and coupling these modes together to form a lattice structure \cite{Yuan2018SyntheticPhotonics}. Synthetic space is not attached to any DOF, it is something that is uniquely convenient to implement in photons because of the photonic engineering which is possible. Based on such an approach, quantum-correlated synthetic crystal is demonstrated by Javid et al.~\cite{Javid2023Chip-scaleSpace} which is based on a coherently controlled broadband QFC produced in a LNOI microresonator incorporating a PPLN region for QFC generation and an electrode for EOM (Fig.~\ref{fig:Tmodes-Hyper-entanglement-Synthetic}F). This approach leverages the time–frequency entanglement within the comb modes to significantly expand the dimensionality of the synthetic space to 400$\times$400 synthetic lattice with electrically controlled tunability.

\section{Applications}

Numerous experiments have explored the generation and distribution of high-dimensional entangled states using on-chip SFWM and SPDC. Future advancements in optical telecommunications involving quantum photonics are anticipated to utilize low-loss optical fiber networks and high-speed photonic interconnect technologies. Consequently, current research in quantum photonics predominantly focuses on scalable and reliable 1550 nm sources, modulators, circuits, and detectors. However, the integration of generation, manipulation, and measurement sections onto a single chip creates new challenges. The manipulation and measurement of high-dimensional entangled states on-chip necessitates both photon sources with high brightness and the integration of various active optical components. Quantum light applications range from quantum communication and computing to imaging and sensing. Below, we highlight recent demonstrations utilizing chip-scale quantum light sources for generating high-dimensional entanglement. \\

To address the maturity and scalability of silicon photonic sources to generate multidimensional quantum entanglement, Wang et.al.~\cite{Wang2018MultidimensionalOptics} demonstrated 16 identical spiral waveguides and over 550 optical components on a single chip.  Their work demonstrates that compact high-dimensionally-entangled photon sources, and certification of randomness and entangled states via Bell inequalities are possible on a fully integrated platform, paving the way for high-dimensional quantum communication. A hardware platform supporting the integration of various quantum information carrier components is required to implement quantum algorithms. 
The number of integrated components on a single chip has seen exponential growth, currently reaching a record of 2,500 components for monolithic integration. Bao et. al.~\cite{Bao2023Very-large-scalePhotonics} integrated an array of 32 spiral SFWM degenerate photon-pair sources to show the generation of genuine multipartite multidimensional quantum entanglement. Interconnects are required for distributed quantum computing regardless of the physical platform that does the quantum computation. Interconnects are also important for quantum networks that feature several nodes for effective entanglement distribution. For both quantum computing and quantum communication, interconnects are necessary for architectural flexibility. Recent advancements demonstrate interconnection between multiple chips, pointing to the feasibility of large-scale practical entanglement distribution. Wang et al.~\cite{Wang2016Chip-to-chipInterconversion} achieved the conversion between path and polarization entanglement across chips by demonstrating chip-to-chip entanglement distribution using two spiral waveguides. The integration of microring resonators and programmable quantum photonic circuits have facilitated chip-to-chip quantum teleportation and entanglement swapping of frequency-encoded quantum states \cite{Llewellyn2020Chip-to-chipSilicon}. Hybrid multiplexing using polarisation- and mode-encoding can also be used for distributing multiple multidimensional entangled states across multiple chips linked by few-mode fibres \cite{Zheng2023MultichipRetrievability}.\\

Quantum key distribution is a major quantum technology that provides unprecedented security of the keys. Because QKD is largely about transmission of keys, QKD is naturally photonic. Increased quantum information capacity can be achieved using qudits. Using integrated sources greatly improves the scalability of QKD, as in \cite{Cabrejo-Ponce2023High-DimensionalDomain} and \cite{Imany201850-GHz-spacedMicroresonator} which used entangled frequency modes which are compatible with telecom optical fibres. Although theoretically secure, implementations of QKD are open to loopholes that undermine security. Measurement-device-independent QKD (MDI-QKD) has been proposed to tackle the imperfections related to the measurement devices \cite{Lo2012Measurement-device-independentDistribution}. \\

\begin{figure} [hbt!]
    \centering
    \includegraphics[width=\textwidth]{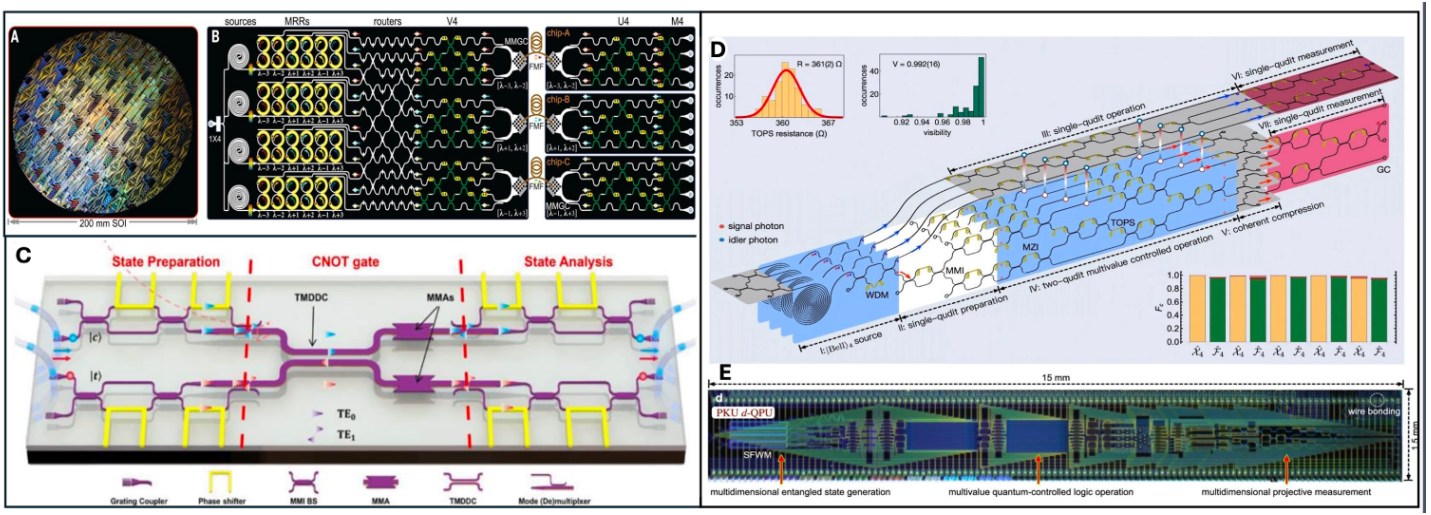}
    \caption{\textbf{Applications of high-dimensional qudit sources on integrated platforms.} \textbf{(A-B)} Demonstration of multichip multidimensional entanglement networks on silicon platform~\cite{Zheng2023MultichipRetrievability}. \textbf{(A)} An image of 200 mm SOI wafer containing 27 copies of quantum networking chips~\cite{Zheng2023MultichipRetrievability}. \textbf{(B)} A completely connected entanglement network comprises one central server chip and various quantum node chips. Photon pairs were generated on the server chip via SFWM process by coherent pumping of four spiral waveguides, producing three pairs of four-dimensional entangled photons. The server chip was connected by three separate few-mode fibres (FMFs) to three node chips (A, B, and C) for implementing multidimensional entanglement distributions across multiple silicon chips~\cite{Zheng2023MultichipRetrievability}
    \textbf{(C)} A multimode waveguide two-qubit CNOT gate circuit demonstrated on silicon photonic chip. The CNOT gate section in the figure represents the multimode section supporting TE$_0$ and TE$_1$ transverse modes for universal transverse mode-encoded quantum operations~\cite{Feng2022TransverseChip}. \textbf{(D-E)} Qudit-based quantum processing unit (d-QPU) on silicon photonic chip~\cite{Chi2022AProcessor}. \textbf{(D)} A schematic of the programmable qudit processor demonstrating the implementation of several algorithms, such as, quantum Fourier transform. A four-level entangled state was generated in an array of four integrated identical SFWM sources for programmable qudit-based quantum computation~\cite{Chi2022AProcessor}. \textbf{(E)} A microscopic image of the d-QPU chip monolithically integrating 4 SFWM sources and more than 400 other photonic components~\cite{Chi2022AProcessor}. Panels reproduced with permission from: \textbf{(A, B)} ref.~\cite{Zheng2023MultichipRetrievability} Springer Nature; \textbf{(C)} ref.~\cite{Feng2022TransverseChip} APS; \textbf{(D, E)} ref.~\cite{Chi2022AProcessor} under a Creative Commons Attribution 4.0 International License (\url{http://creativecommons.org/licenses/by/4.0/}).}
    \label{fig:Applications}
\end{figure}

Quantum computation is another application for which a photonic system is attractive. Quantum computation is possible using just linear optics \cite{Knill2001AOptics}. Post-selection is used to prepare entangled states, which is possible with gate teleportation, but is very resource-intensive. While deterministic gate requirements pose practical challenges, the measurement-based quantum computing (MBQC) model offers a more resource-efficient alternative. Any circuit-based quantum computation can be mapped to MBQC \cite{Raussendorf2001AComputer,Nielsen2004OpticalStates,Browne2005Resource-efficientComputation,Carolan2015UniversalOptics}. Reimer et.al.~\cite{Reimer2019High-dimensionalStates} utilized SFWM within a microring resonator to implement three-level, four-partite cluster states formed by two photons in the time and frequency domain for high-dimensional one-way quantum operations. Operating on transverse modes on a silicon photonic chip, \cite{Feng2022TransverseChip} demonstrated a two-qubit quantum gate pointing to the possibility of universal transverse mode-encoded quantum operations Fig.~\ref{fig:Applications}C. Operating on frequency modes, Lu et.al.~\cite{Lu2019AQubits} designed a CNOT gate for a QFC coming from a PPLN waveguide. Building on \cite{Lu2019AQubits} more modes were used in \cite{Imany2019High-dimensionalSpaces} to encode qudits in time and frequency DoFs, reporting deterministic two-qudit gates on the chip silicon nitride microresonator. A programmable qudit-based quantum processor was demonstrated by Chi et.al~\cite{Chi2022AProcessor} on a silicon-photonic integrated circuits. This implementation allowed more than one million configurations that demonstrate high-fidelity quantum state preparation, operation, and measurement, benchmarked via different quantum algorithms.\\

The goal for many quantum computing systems is to demonstrate a quantum computational advantage---some problems are more efficiently solved by a quantum computer than with a classical computer. It is expected that the competition between the best classical strategies and what can be achieved using a quantum computer will continue. Boson sampling is one class of problems for which we have seen such a competition emerge. Boson sampling refers to sampling probability distributions of the output when an $n-$ boson state undergoes linear scattering---a universal quantum computer is not necessary. Because photons interact only linearly in a linear-optical quantum network, boson sampling is suited to photonic systems and there have been several photonic experiments \cite{Broome2013PhotonicCircuit,Brod2019PhotonicReview}. Wang et.al.~\cite{Wang2017High-efficiencySampling} implemented high performance multiphoton boson sampling with quantum dot–micropillar single-photon sources. The setup validated boson sampling for three-, four- and five-photons, achieving 4.96 kHz, 151 Hz and 4 Hz sampling rates, respectively. Another experiment by Wang et.al.~\cite{Wang2019BosonSpace} scaled up the boson sampling with 20 photons injected into a 60-mode interferometer, with the output Hilbert space reaching 3.7 $\times$ 10$^{14}$. On chip, Paesani et.al.~\cite{Paesani2019GenerationChip} reported the generation of quantum states of light with up to eight photons, implementing the standard, scattershot, and Gaussian boson sampling protocols in the same silicon chip. The advancements in wafer-scale fabrication processes are driving a significant shift, with large-scale circuits beginning to transition towards utility-scale devices. Such progression highlights the growing feasibility and potential of integrating high-dimensional entangled photon sources into practical quantum technologies, paving the way for broader applications in quantum communication and computing.

\section{Outlook and Conclusion}

Photonic integration presents a robust strategy for the miniaturization and scaling of current state-of-the-art quantum technologies \cite{Alexander2024AComputing, Luo2023RecentInternet}. This integration is critical for achieving the fault tolerance and error correction necessary for the realization of practical and scalable quantum computing systems \cite{Alexander2024AComputing}. The wide interest in photonic quantum computing, from both academia and industry, will fuel future advances in both photonic hardware and software (e.g. algorithms and benchmarking that are uniquely suitable to photons) \cite{Romero2024PhotonicComputing, McMahon2023TheComputing,Alexander2024AComputing}. Regardless of the photonic quantum technology, entangled photon sources are important. Having these sources miniaturised on a chip is beneficial for real-world applications.  The source of entangled photons is just one (albeit important) component. To achieve widespread practical applications, components for processing and detecting the entangled quantum states should be integrated on chip too. \\

This article summarised recent developments in generating entangled qudits on-chip. These developments heavily rely on integration. No single material platform excels in all metrics necessary for quantum applications. Achieving full system integration on a single chip involves combining photon sources, single-photon detectors, lasers, and modulators into a unified platform. This integration is essential for creating robust, compact, scalable, and high-performance quantum photonic circuits. The mature silicon-on-insulator CMOS fabrication techniques enabled the integration of a large number of photonic components monolithically on a single silicon chip \cite{Wang2018MultidimensionalOptics}. However, the two-photon absorption on increasing the pump power makes it difficult to enhance the pair generation rate in Si. Conversely, the negligible two-photon absorption in SiN and fast on-chip modulators in LN favouring the next generation of integrated quantum photonics. These factors drive the adoption of hybrid and heterogeneous integration strategies \cite{Elshaari2020HybridCircuits} that leverage the strengths of each platform to achieve optimal performance across the various components necessary for quantum computing \cite{Mahmudlu2023FullyGeneration,Liu2024ALaser,Najafi2015On-chipDetectors,Wang2018IntegratedVoltages,Alexander2024AComputing}. \\

\begin{figure} [hbt!]
    \centering
    \includegraphics[width=\textwidth]{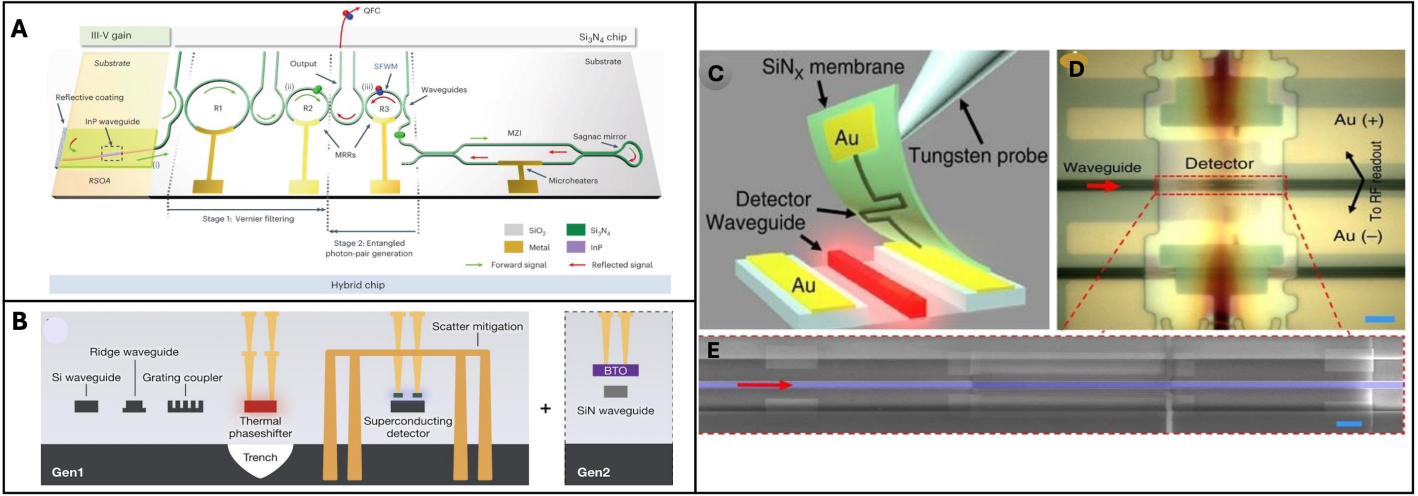}
    \caption{\textbf{Hybrid integration on integrated platforms.} \textbf{(A)} A complete integrated hybrid quantum light source comprising a laser cavity (on InP platform), tunable noise suppression filter, and an entangled photon pair generating source. High-dimensional frequency-bin entangled qudits were generated via SFWM in a SiN microring~\cite{Mahmudlu2023FullyGeneration}. \textbf{(B)} PsiQuantum hybrid integrated photonics platform for quantum computation. They have demonstrated a photon source (SiN photonic waveguides), single-photon detectors (Niobium Nitride (NbN) superconducting layer), and waveguide-integrated BTO EO phase shifters on one quantum photonic chip \cite{Alexander2024AComputing}. \textbf{(C-E)} Implementation of on-chip single photon detectors (SNSPDs)~\cite{Najafi2015On-chipDetectors}. \textbf{(C)} Method to transfer detectors onto a Si waveguide. Each of the detectors used in the experiment was fabricated by the deposition of NbN film on top of the SiN layer~\cite{Najafi2015On-chipDetectors}. \textbf{(D)} Optical image of single photon detector (SNSPD) integrated with a Si waveguide \cite{Najafi2015On-chipDetectors}. \textbf{(D)} SEM image of the detector \cite{Najafi2015On-chipDetectors}. Panels reproduced with permission from: \textbf{(A)} ref.~\cite{Mahmudlu2023FullyGeneration}  under a Creative Commons Attribution 4.0 International License (\url{http://creativecommons.org/licenses/by/4.0/}); \textbf{(B)} ref.~\cite{Alexander2024AComputing} under \url{http://arxiv.org/licenses/nonexclusive-distrib/1.0/}; \textbf{(C, D, E)} ref.~\cite{Najafi2015On-chipDetectors} under a Creative Commons Attribution 4.0 International License (\url{http://creativecommons.org/licenses/by/4.0/}).}
    \label{fig:Hybrid}
\end{figure}

There is another technique---inverse design---which has significantly mitigated fabrication and technical challenges in silicon nanophotonics by utilizing computational approaches to discover optimal optical structures based on desired functional characteristics \cite{Molesky2018InverseNanophotonics}. Inverse design employs algorithmic techniques, such as genetic and gradient-based algorithms, to optimize structures over a vast design space, enabling the creation of devices with superior performance metrics \cite{Molesky2018InverseNanophotonics}. Inverse-designed passive components, such as mode multiplexers and beam splitters for silicon photonic circuits, are already established. Expanding inverse design to include active devices, such as modulators and lasers—which frequently limit performance in optical systems—would significantly enhance monolithic integration and advance the capabilities of integrated quantum photonics \cite{Molesky2018InverseNanophotonics}. \\

Over recent decades, the field has overcome numerous technological and manufacturing challenges. This rapid progression has fueled immense anticipation for the realization of large-scale integrated nonlinear photonics in quantum computing \cite{Alexander2024AComputing}, quantum communication \cite{Labonte2024IntegratedMetrology}, neuromorphic computing \cite{Farmakidis2024IntegratedChallenges}, and quantum optics \cite{Dutt2024NonlinearMaterials}. In the upcoming years, qudit-based technology needs further research and development to precisely control integrated sources along with other photonic components to showcase quantum operations on a single photonic device.


\bibliographystyle{elsarticle-num}
\bibliography{references-final.bib}

\end{document}